\documentclass[ip,a4paper,twocolumn]{jpsj3}
\usepackage{txfonts}
\usepackage{bm}

\title{Superconductivity in Ce- and U-based "122" heavy-fermion compounds}

\author{$^1$Oliver Stockert, $^{1,2}$Stefan Kirchner, $^1$Frank Steglich, $^3$Qimiao Si
}
\inst{\address{$^1$Max-Planck-Institut f\"ur Chemische Physik fester Stoffe, Dresden, Germany}\\
\address{$^2$Max-Planck-Institut f\"ur Physik komplexer Systeme, Dresden, Germany}\\
\address{$^3$Dept. of Physics and Astronomy, Rice University, Houston, Texas, USA}
}
\abst{This review discusses the heavy-fermion superconductivity in Ce- and U-based compounds crystallizing in the body-centered tetragonal ThCr$_2$Si$_2$ structure. Special attention will be paid to the theoretical background of these systems which are located close to a magnetic instability.}

\kword{heavy-fermion superconductivity, cerium, uranium, quantum phase transition}

\begin{document}
\maketitle

\section{Introduction}
Unconventional superconductivity most commonly occurs in the vicinity of magnetic order.  It is to be contrasted with BCS superconductivity, which is driven by electron-phonon interactions and to which magnetism is detrimental~\cite{Tinkham}. The magnetism in these unconventional superconductors signals the importance of electron-electron repulsion, which gives rise to several related effects.  In addition to magnetic order, the electron correlations promote the localization of charge carriers, and can give rise to unusual normal states. Heavy fermion superconductors represent prototype systems for the interplay between such correlation effects and unconventional superconductivity. 

The microscopic physics of heavy fermion metals is described in terms of the competition between the Kondo effect and the Ruderman-Kittel-Kasuya-Yosida (RKKY) interaction of the f electrons~\cite{hewson93,stewart01,loehneysen07a,gegenwart08}. The competing interactions give rise to a variety of phases, including magnetic order and paramagnetic heavy Fermi liquid. Superconductivity usually occurs near the transition between the corresponding ground states. 

The $T=0$ transition between the competing ground states can be either of first order or continuous. In the latter case, the transition specifies a quantum critical point~\cite{coleman05,Si.10}. The associated quantum fluctuations determine the behavior at low energies, giving rise to non-Fermi-liquid properties~\cite{maple94,aronson95,stewart01,loehneysen07a}. Many heavy-fermion materials have been shown to display such quantum critical behavior, which is typically revealed through thermodynamic and transport measurements. For instance, the electrical resisivity typically exhibits a power-law dependence on temperature with unusual exponents, $\Delta \rho = \rho(T)-\rho_0 \propto T^n$, $n <  2$ (in contrast to $n =  2$ for the Fermi-liquid behavior).

\begin{figure}[t]
\centering
\includegraphics[width=\linewidth,clip]{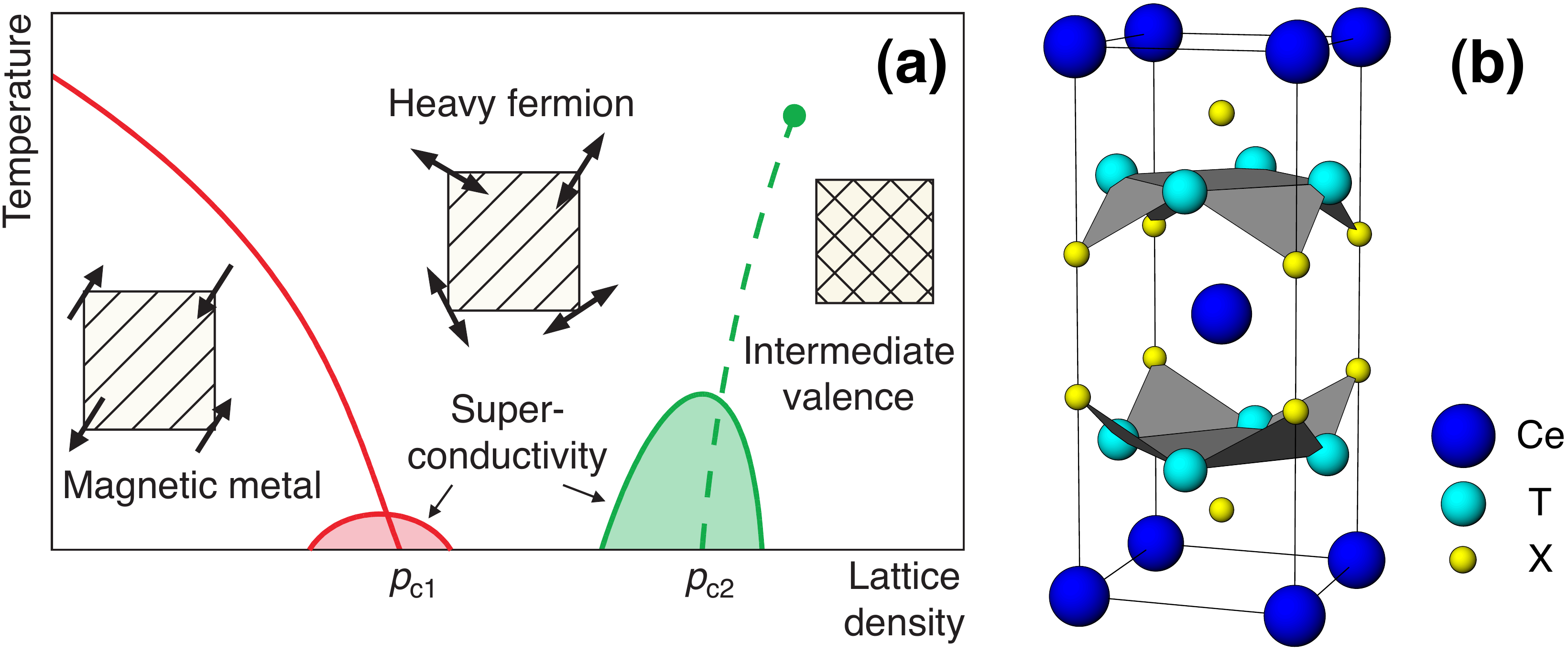}
\caption{(Color online) (a) Schematic phase diagram near quantum phase transitions in heavy-fermion systems~\cite{yuan03}. (b) Body-centered tetragonal crystal structure of Ce-based CeT$_2$X$_2$ (T =  Cu, Rh, Ni, Pd; X = Si, Ge) heavy-fermion compounds.\label{fig1}}
\end{figure}

This is illustrated by the schematic phase diagram\cite{yuan03}, Fig.\,\ref{fig1}a, for CeCu$_2$(Si$_{1-x}$Ge$_{x}$)$_2$. Here, the control parameter that tunes between the ground states is hydrostatic pressure (and hence the lattice density). Superconductivity shows up around the magnetic instability at $p_{c1}$  where antiferromagnetic order is fully suppressed.
Fig.\,\ref{fig1}a also contains a second dome of superconductivity near $p_{c2}$, which is connected with a first-order valence transition. This illustrates that the ground states involved in the transitions can go beyond magnetism. There are also cases where the underlying order remains mysterious, often termed ``hidden'' order.

In this article, we review the physical properties  of the ``122'' heavy-fermion superconductors. At present, 
this family is  formed by the Ce-based compounds, CePd$_{2}$Si$_{2}$, CeNi$_2$Ge$_2$, CeRh$_2$Si$_2$, CeCu$_2$Ge$_2$,
and CeCu$_2$Si$_2$ and the U-based hidden order compound URu$_2$Si$_2$.
The 122 compounds crystallize in the body-centered
tetragonal ThCr$_2$Si$_2$  structure (space group I4/mmm ) shown
in Fig. 1b. Within this structure the magnetic 4f ions occupy
the corners and the center of the unit cell. Since the
next-nearest neighbor distances of the 4f ions within the
basal plane and from the corner to the center of the unit cell
are almost identical, the interactions in these compounds are
truly three-dimensional, in contrast to, e.g., the layered perovskite
high-temperature cuprate superconductors. This
lattice implies that magnetic frustration may
exist in the 122
heavy-fermion systems~\cite{grosche01}. CePd$_2$Si$_2$, CeRh$_2$Si$_2$  and CeCu$_2$Ge$_2$
 are well-ordered antiferromagnets with N\'eel temperatures of
several Kelvin, and magnetic order vanishes only upon applying
a pressure $p \gtrsim 1$\,GPa where superconductivity appears.
In contrast, CeCu$_2$Si$_2$ and CeNi$_2$Ge$_2$ are located at or very
close to a magnetic instability already at ambient conditions,
and a subtle interplay between magnetism and superconductivity
occurs. In general, superconductivity
in the 122 compounds is observed only in a narrow range
around the zero temperature transitions. However, although magnetic
and valence instabilities are disconnected, they might be close
enough to each other resulting in one extended superconducting
regime as seen in both CeCu$_2$Si$_2$ and CeCu$_2$Ge$_2$. The excited
crystalline electric field (CEF) levels of all above mentioned
compounds are well above 1 meV corresponding to  $\approx 10$\,K \cite{severing89,abe98}. 
\begin{table*}[t]
\begin{center}
\begin{tabular}{|c|c|c|c|c|c|c|c|c|}
\hline
Compound& $T_{\rm N}(p = 0)$& $T_{\rm K}(p = 0)$& $\gamma (p=0,T\rightarrow0)$& propagation& $\mu_{ord}$& $p_c$& $T_c(p = p_c)$& $n$ in $\Delta\rho\propto T^n$\\
 & [K]& [K]& [J/molK$^2$]& vector $\bm Q$&  [$\mu_{\rm B}$]& [GPa]& [K]&  at $p_c$\\ 
\hline
\hline
CePd$_2$Si$_2$& 10.2 \cite{grier84}& 10 \cite{severing89a}& 0.25 \cite{steeman88}& $(1/2~1/2~0)$\cite{grier84,steeman88}& 0.66 \cite{grier84,steeman88}& 2.8 \cite{grosche96,mathur98}& 0.43 \cite{grosche96,mathur98}& 1.2 \cite{grosche96,mathur98}\\
\hline
CeNi$_2$Ge$_2$& --& 30 \cite{knopp88}& $0.35$ \cite{knopp88}& $(1/2~1/2~0)$\cite{kadowaki03}& --& 0& 0.2 \cite{gegenwart99,grosche00,braithwaite00}& 1.1--1.5 \cite{gegenwart99,grosche00}\\
 & & & & $(0~0~3/4)$ \cite{kadowaki03}& & & &\\
 & & & & $(0.23~0.23~0.5)$ \cite{fak00}& & & &\\
\hline
CeRh$_2$Si$_2$&  36 \cite{godart83,settai97}& $30-100$ \cite{severing89a,kawasaki98}& 0.023 \cite{graf97}& $(1/2~1/2~0)$& $1.3-1.4$ \cite{settai97}& 1.06 \cite{kawarazaki00,movshovich96,grosche97}& 0.42 \cite{araki02}& 2 \cite{grosche97}\\
& &&  & $(1/2~1/2~1/2)$ \cite{grier84,kawarazaki00}& & & & \\
\hline
\hline
CeCu$_2$Ge$_2$& 4.15 \cite{knopp89,krimmel97}& 4 \cite{knopp87}& 0.1 \cite{boer87}& $(0.28~0.28~0.54)$ \cite{krimmel97}& 1 \cite{krimmel97}& 9.4 \cite{jaccard92,jaccard99}& 0.6 \cite{jaccard99}& 2 \cite{jaccard99}\\
\hline
CeCu$_2$Si$_2$& 0.8 \cite{nakamura88,uemura88,uemura89}& 10 \cite{severing89a}& 1 \cite{steglich79}& $(0.215~0.215~0.54)$ \cite{stockert04}& 0.1 \cite{stockert04}& 0& 0.6 \cite{steglich79}& 1-1.5 \cite{yuan03,sparn98}\\
\hline
\hline
URu$_2$Si$_2$& 17.5 ($T_0$) \cite{schlabitz86,palstra85,maple86}& 60 \cite{kohara86}& 0.07 \cite{steglich79}& $(1~0~0)$ \cite{broholm87}& 0.03 \cite{broholm87}& 0& 1.5 \cite{schlabitz84,schlabitz86}& ?\\
\hline
\end{tabular}
\end{center}
\caption[]{Thermodynamic and transport properties of Ce-based "122" heavy-fermion superconductors and URu$_2$Si$_2$ ($T_{\rm N}(p = 0)$ and $T_{\rm K}(p = 0)$: N\'eel temperature and Kondo temperature at ambient pressure; $\gamma(p=0,T \rightarrow 0)$: Sommerfeld coefficient of the specific heat at ambient pressure for $T \rightarrow 0$; $\bm Q$ and $\mu_{ord}$: propagation vector and ordered moment of the magnetic order; $p_c$: critical hydrostatic pressure to fully suppress magnetic order; $T_c(p=p_c)$: superconducting transition temperature at $p = p_c$; $n$ at $p_c$: exponent in the temperature dependence of the electrical resistivity at $p_c$, $\Delta\rho \propto T^n$). The precise meaning for multiple entries in the column 'propagation vector' for CeNi$_2$Ge$_2$ and CeRh$_2$Si$_2$ is given in the text. Displayed Kondo temperatures $T_{\rm K}$ have been determined mainly by microscopic measurements (e.g., quasielastic neutron scattering).\label{table1}}  
\end{table*}
Therefore, their low-temperature magnetic behavior
is entirely determined by the spin dynamics within the
CEF ground state doublet of the cerium 4f moment \cite{grier84,severing89a,loidl92}. Table \ref{table1}
summarizes key properties of the relatively small number of Ce-based 122 heavy-fermion
compounds and URu$_2$Si$_2$ which become superconducting. We will
discuss them in more detail in the following.

\section{Superconductivity near magnetic quantum critical points: CePd$_2$Si$_2$ and CeNi$_2$Ge$_2$}

\begin{figure}[tb]
\centering
\includegraphics[width=\linewidth,clip]{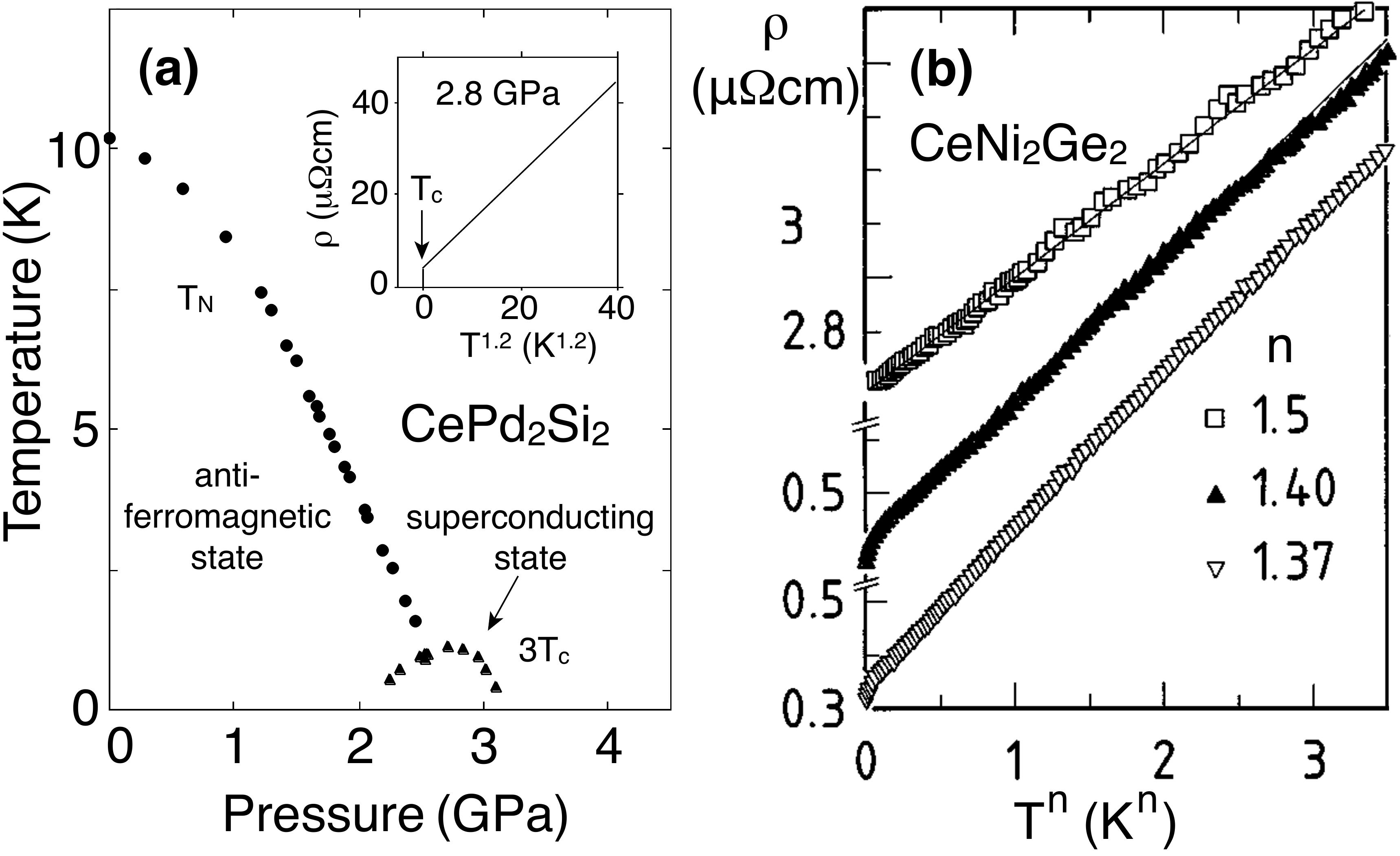}
\caption{(a) Temperature-pressure phase diagram of CePd$_2$Si$_2$. Superconductivity appears only in a narrow pressure range where the N\'eel temperature goes to zero. Inset: the normal-state resistivity along the $a$-axis varies as $\rho \propto T^{1.2}$ over almost two decades in temperature \cite{mathur98}. (b) Electrical resistivity of CeNi$_2$Ge$_2$ at low temperatures for three samples with different purities plotted as $\rho$ vs. $T^n$ \cite{gegenwart99}.\label{fig2}
}
\end{figure}

CePd$_2$Si$_2$ is one of the first compounds in which magnetically mediated superconductivity around a pressure-induced quantum critical point was suggested \cite{grosche96,mathur98}. It orders antiferromagnetically below $T_{\rm N} \approx 10.2$\,K in a simple commensurate structure with a propagation vector $\bm Q = (1/2~1/2~0)$ \cite{grier84,steeman88}. 
The ordering temperature is quite high, of the same size as the Kondo temperature (cf. Table\,\ref{table1}). The ordered moment reaches a value of only $0.66\,\mu_{\rm B}$ \cite{grier84,steeman88} corresponding to an enhanced specific heat coefficient $\gamma \approx 0.25$\,J/molK$^2$ \cite{steeman88}. As displayed in Fig.\,\ref{fig2}(a) applying hydrostatic pressure successively reduces the antiferromagnetic order in CePd$_2$Si$_2$, and superconductivity appears in a narrow range around the critical pressure $p_c \approx 2.8$\,GPa where magnetism is fully suppressed \cite{mathur98}. The superconducting transition temperature reaches a maximum value of $T_c = 0.43$\,K. The electrical resistivity displays clear non-Fermi-liquid behavior in the normal state above the superconducting dome around $p_c$, i.e., $\Delta\rho\propto T^{1.2}$ over almost 2 decades in temperature as seen in the inset of Fig.\,\ref{fig2}(a). The commensurate magnetic structure remains unchanged under pressure \cite{kernavanois05} as revealed by single crystal neutron diffraction up to $2.45$\,GPa, i.e., close to $p_c$. 
According to these neutron scattering experiments the ordered moment decreases linearly with smaller N\'eel temperature which was taken as a strong support for an itinerant spin-density-wave scenario at the quantum critical point \cite{kernavanois05}. 

Among the Ce-based 122 compounds CeNi$_2$Ge$_2$ takes a special role in the respect that this heavy-fermion compound ($\gamma \approx 0.35$\,J/molK$^2$, $T_{\rm K} \approx 30$\,K \cite{knopp88}) is not magnetically ordered, but is located already at ambient pressure near a magnetic instability of spin-density-wave type \cite{kuechler07}. This vicinity to magnetic order is manifested by alloying experiments indicating that already a small amount of doping on the Ni site is sufficient to induce antiferromagnetic order. Via isoelectronic substitution antiferromagnetism appears in Ce(Ni$_{1-x}$Pd$_x$)$_2$Ge$_2$ above a critical Pd concentration $x \approx 0.065 - 0.1$ \cite{fukuhara98,knebel99}, while in Ce(Ni$_{1-x}$Cu$_x$)$_2$Ge$_2$ long-range antiferromagnetism occurs for a Cu content $x> 0.25$ \cite{loidl92}.
In thermodynamic and transport properties CeNi$_2$Ge$_2$ exhibits pronounced non-Fermi-liquid behavior, e.g., the electrical resistivity varies with temperature as $\Delta\rho\propto T^{n}, n = 1.1-1.5$ \cite{gegenwart99,grosche00}. The exponent $n$ is lower for samples of higher purity as expected theoretically \cite{rosch99} and as shown in Fig.\,\ref{fig2}(b) where the electrical resistivity $\rho$ is plotted versus $T^n$ \cite{gegenwart99}. The purer the samples, the larger are the deviations from Fermi-liquid behavior. While Pd-doped samples at the critical Pd concentration obey a square-root temperature dependence of the specific-heat coefficient $\gamma = C/T$ \cite{wang11}, i.e., $C/T = \gamma_0 - a T^{1/2}$, non-Fermi-liquid behavior with a logarithmic divergence in $C/T$ is present for pure CeNi$_2$Ge$_2$ \cite{gegenwart99},  $C/T \propto \ln T/T_0$.
Moreover, in highest-purity CeNi$_2$Ge$_2$ samples incipient superconductivity is found below $T_c \approx 0.2$\,K \cite{steglich97,gegenwart99,grosche00,braithwaite00}. 
It is worth mentioning that in addition to the superconducting dome around the magnetic instability at ambient pressure a second extended superconducting regime has been observed in resistivity measurements at higher pressure between $p \approx 1.5$ and $6.5$\,GPa with a maximum $T_c \approx 0.4$\,K \cite{grosche00,braithwaite00}.
Neutron scattering experiments on pure CeNi$_2$Ge$_2$ so far have focused on the magnetic response at ambient pressure.
Incommensurate spin fluctuations have been detected with a wave vector $\bm Q_1 = (0.23~0.23~0.5)$ and a characteristic energy of $4$\,meV \cite{fak00}. In addition, magnetic correlations occur at low temperatures at the commensurate wave vectors $\bm Q_2 = (1/2~1/2~0)$ and $\bm Q_3 = (0~0~3/4)$ \cite{kadowaki03}, whose energy scale is quite small with $0.6$\,meV, as expected for a compound close to magnetic order.

\section{Superconductivity near first order transitions: CeRh$_2$Si$_2$ and CeCu$_2$Ge$_2$}

\begin{figure}[b]
\centering
\includegraphics[width=\linewidth,clip]{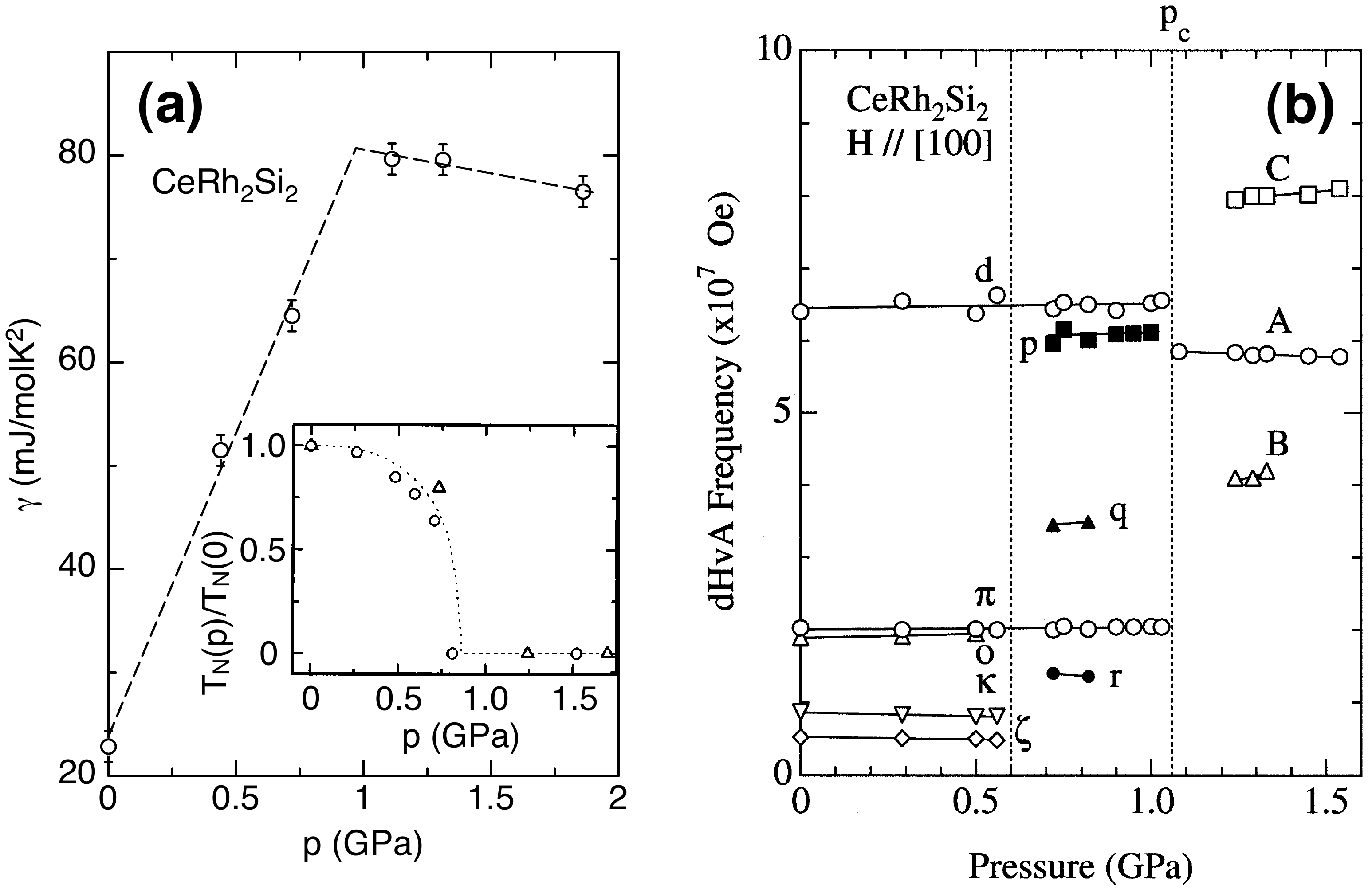}
\caption{(a) The coefficient $\gamma$ of the linear-in $T$ specific heat of CeRh$_2$Si$_2$ as a function of hydrostatic pressure $p$. Inset: Normalized  N\'eel temperature $T_{\rm N}(p)/T_{\rm N}(0)$ versus pressure $p$. All lines are just guides to the eyes \cite{graf97}. (b) Pressure dependence of the de Haas-van Alphen (dHvA) frequencies in CeRh$_2$Si$_2$ for magnetic field applied along the $a$ axis \cite{araki01,araki02a}. The data were taken at temperatures $T = 102 - 160$\,mK. \label{fig3}}
\end{figure}

CeRh$_2$Si$_2$ exhibits antiferromagnetic order with a quite high N\'eel temperature $T_{\rm N} = 36$\,K at ambient pressure \cite{godart83,settai97}, which has to be compared to a correspondingly high Kondo temperature $T_{\rm K}$ \cite{severing89a,kawasaki98} (cf. Table\,\ref{table1}). As revealed by powder neutron diffraction the antiferromagnetic structure is commensurate with a propagation vector $\bm Q_1 = (1/2~1/2~0)$ just below $T_{\rm N}$, followed by a second magnetically ordered phase for $T < 25$\,K and characterized by an additional propagation vector $\bm Q_2 = (1/2~1/2~1/2)$ \cite{grier84}. The ground state magnetic structure has been determined as a 4$q$ structure with the two propagation vectors $\bm Q_1$ and $\bm Q_2$ \cite{kawarazaki00}. In both phases the ordered magnetic moments of the two magnetic sublattices are aligned along $[001]$. Their magnitude attains $\mu_{ord} \approx 1.3-1.4\,\mu_{\rm B}$ at low temperatures, values quite close to the expected value for the CEF ground state doublet. 
Upon applying hydrostatic pressure the ordered magnetic moment and the N\'eel temperature $T_{\rm N}$ of CeRh$_2$Si$_2$ are successively suppressed, see inset of Fig.\,\ref{fig3}(a) \cite{kawarazaki00,graf97}. Despite the high ordering temperature of $T_{\rm N} = 36$\,K only a moderate pressure $p_c \approx 1.06$\,GPa is needed to fully suppress antiferromagnetism, and bulk superconductivity appears in a very narrow pressure range around the critical pressure $p_c$ \cite{movshovich96,grosche97,kawarazaki00}. The superconducting $T_c$ reaches a maximum at $p_c$ with $T_c \approx 0.42$\,K \cite{araki02}. It should be noted that antiferromagnetism vanishes in a first order phase transition at $p_c$ as recently proven by clear volume discontinuities in thermal expansion measurements under pressure around $p_c$ \cite{villaume08,boursier08}. Moreover, at the magnetic instability Fermi-liquid behavior is observed. For instance, the electrical resistivity displays Fermi-liquid behavior even around $p_c$ with a $\Delta \rho = A T^2$ temperature dependence \cite{grosche97}. However, the $A$ coefficient of the electrical resistivity shows a pronounced maximum at the critical pressure $p_c$ \cite{araki02,boursier08}. At the same time, the temperature-independent specific heat coefficient $\gamma$ rises steeply from ambient pressure, where the mass enhancement is small ($\gamma = 23$\,mJ/molK$^2$) up to $p_c$ where $\gamma \approx 80$\,J/molK$^2$ is found (cf. Fig.\,\ref{fig3}(a)) \cite{graf97}. Beyond $p_c$ only a slight decrease in $\gamma$ is observed.  As displayed in Fig.\,\ref{fig3}(b), de Haas-van Alphen measurements under pressure indicate an almost unchanged Fermi surface up to $1$\,GPa (some branches appear/disappear around $0.6$\,GPa due to a change in magnetic structure from the 4$q$ to a single $\bm Q_1$ structure), but point to a sudden change of the Fermi surface topology at $p_c$ \cite{araki01,araki02a}. This abrupt change in the Fermi surface topology at $p_c$ is interpreted as a transition from localized 4f electrons at low pressure to itinerant 4f electrons above $p_c$.

\begin{figure}[t]
\centering
\includegraphics[width=0.85\linewidth,clip]{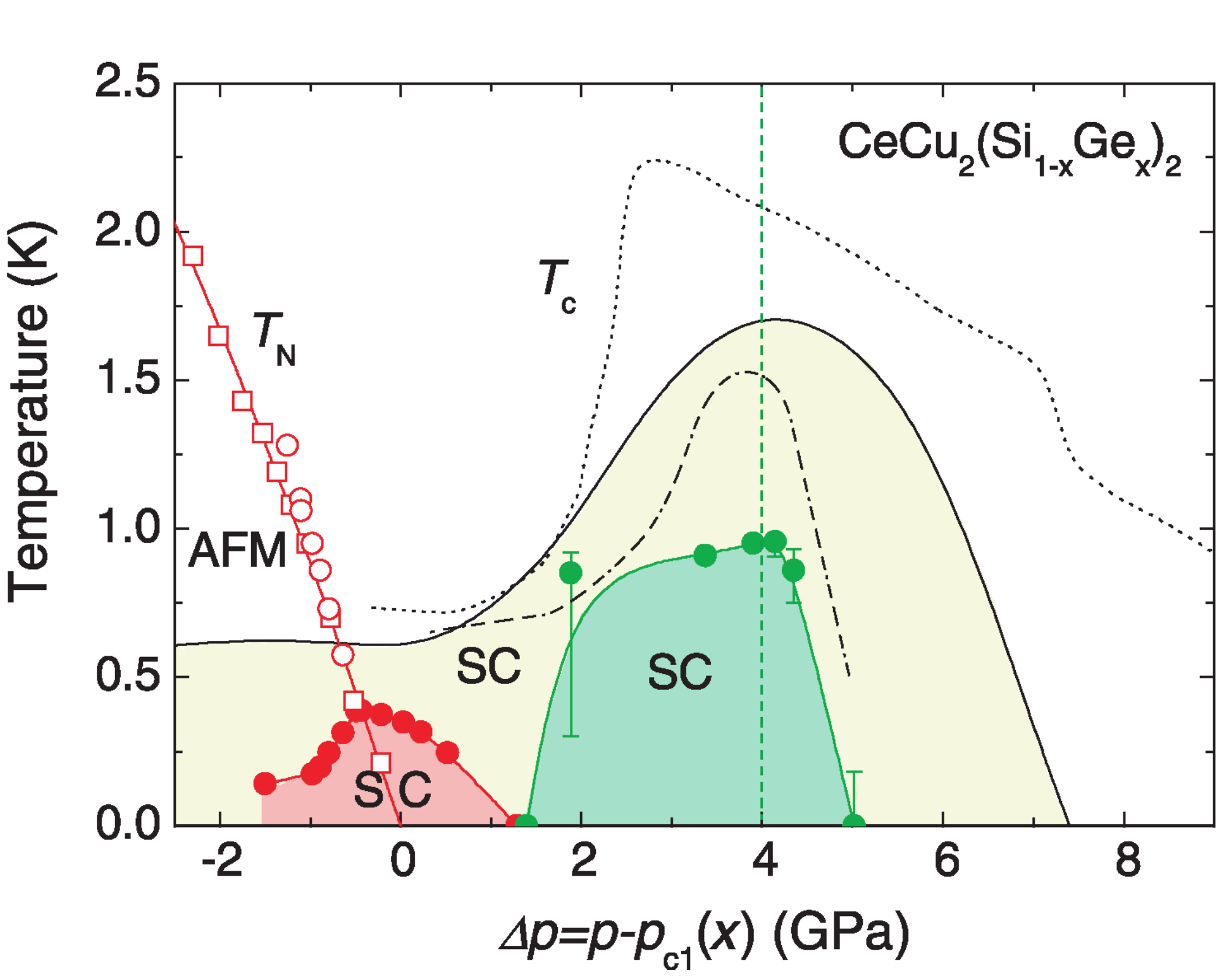}
\caption{(Color online) Temperature-pressure phase diagram of CeCu$_2$(Si$_{1-x}$Ge$_x$)$_2$ displaying the antiferromagnetic and superconducting phase transitions $T_{\rm N}$ and $T_c$ in dependence of the relative pressure $\Delta p = p - p_{c1}$ with $p_{c1}$ being the critical pressure to suppress antiferromagnetic order ($x = 0.1$: circles; $x = 0.25$: squares) \cite{yuan03}. While for pure CeCu$_2$Si$_2$ (dashed and dotted lines) and CeCu$_2$Ge$_2$ (solid line) an extended pressure regime is observed where superconductivity occurs, two distinct superconducting domes are present for $10$\% Ge doped CeCu$_2$Si$_2$ (circles). \label{fig4}}
\end{figure}

CeCu$_2$Ge$_2$ can be regarded as the well magnetically ordered sister compound to the heavy-fermion superconductor CeCu$_2$Si$_2$. Below the N\'eel temperature of $T_{\rm N} = 4.15$\,K, CeCu$_2$Ge$_2$ orders in an incommensurate magnetic structure with a wave vector $\bm Q = (0.28~0.28~0.54)$ and an ordered moment of roughly $1\,\mu_{\rm B}$ \cite{knopp89,krimmel97}. With a Kondo temperature similar to the ordering temperature \cite{knopp87} and a low Sommerfeld coefficient of the specific heat \cite{boer87} (see Table\,\ref{table1}) CeCu$_2$Ge$_2$ is believed to be a local-moment antiferromagnet. However, theoretical calculations indicate that a full local-moment picture cannot account for the observed incommensurate propagation vector, but a certain degree of itineracy is needed \cite{zwicknagl07}. Under hydrostatic pressure magnetic order is suppressed discontinuously and superconductivity occurs \cite{jaccard92}. At the critical pressure $p_c \approx 9.4$\,GPa necessary  to fully suppress antiferromagnetism the superconducting $T_c$ attains a value $T_c \approx 0.6$\,K \cite{jaccard99}. In contrast to other systems which exhibit a superconducting dome just around $p_c$, the superconducting $T_c$ in CeCu$_2$Ge$_2$ remains almost constant with further increasing pressure up to $p \approx 13$\,GPa, as seen in Fig.\,\ref{fig4}. For even higher pressures $T_c$ shows a strong increase with a maximum of $T_c \approx 2$\,K at $p \approx 16$\,GPa. Only beyond $16$\,GPa superconductivity is suppressed again \cite{jaccard99}. At $p_c$, i.e., at the first-order antiferromagnetic quantum phase transition, Fermi-liquid properties are found in the electrical resistivity, i.e., $\Delta\rho = A T^n$, $n = 2$ with just a broad maximum in $A$, while at higher pressure the temperature exponent $n$ decreases and reaches almost $n = 1$ around $p = 15$\,GPa.

\section{Superconductivity in CeCu$_2$Si$_2$}

We now turn to an in-depth discussion of the superconducting properties of CeCu$_2$Si$_2$.
As for CeNi$_2$Ge$_2$, CeCu$_2$Si$_2$ is located already at ambient conditions at a magnetic instability, with the appearance of unconventional superconductivity around the quantum critical point. Thus, magnetically ordered and superconducting ground states are experimentally easily accessible either at ambient or very low hydrostatic pressures or by small Ge substitution on the Si site. CeCu$_2$Si$_2$ was the first unconventional superconductor with $T_c \approx 0.6$\,K, discovered in 1979 \cite{steglich79}. Ten years later it was realized by NQR and $\mu$SR measurements \cite{nakamura88,uemura88,uemura89} that a magnetically ordered state does exist in CeCu$_2$Si$_2$ below $T_{\rm N} \approx 0.8$\,K. Since at that time no sufficiently large single crystals were available to perform neutron-scattering experiments, the nature of the magnetic order remained unidentified. From a tiny (0.5\%) increase in the electrical resistivity below the magnetic ordering temperature \cite{gegenwart98} it was suggested that CeCu$_2$Si$_2$ orders in a small moment spin-density wave, with the opening of a spin gap. The ground state in CeCu$_2$Si$_2$ depends very sensitively on the actual composition within the narrow homogeneity range in which this tetragonal heavy-fermion compound forms. After detailed investigations with different compositions \cite{steglich96} it was concluded that slight Si excess leads to antiferromagnetic order (A-type), while in contrast Si deficiency or Cu excess results in purely superconducting samples (S-type). In samples crystallizing with the nominal 1\,:\,2\,:\,2 stoichiometry both superconductivity and antiferromagnetism are observed. In these A/S-type samples there is a first-order transition from the antiferromagnetic phase at higher temperature to superconductivity at lower temperature. The two orders exclude each other on a microscopic scale \cite{stockert06b}.

As already mentioned, CeCu$_2$Si$_2$ can be easily tuned through the quantum critical point by external hydrostatic pressure as depicted in Fig.\,\ref{fig1}. An initial magnetically ordered ground state is successively suppressed and superconductivity appears around the quantum critical point. The superconducting ground state extends to a pressure as high as $10$\,GPa and is not restricted to the close vicinity of the quantum critical point. Studying the normal state properties at the quantum critical point, i.e., when applying an overcritical magnetic field to suppress superconductivity, one observes in S-type CeCu$_2$Si$_2$ non-Fermi-liquid behavior in the heat capacity and electrical resistivity, which is indicative of a three-dimensional spin-density-wave type antiferromagnetic instability. This was inferred from the observation that the Sommerfeld coefficient of the heat capacity depends on temperature as $C/T \propto \gamma_0 -\alpha\sqrt{T}$ \cite{sparn98} and the electrical resistivity varies with temperature as $\Delta\rho \propto T^{n}$, $n=1 \ldots 1.5$ \cite{sparn98,yuan03}.
A recent pressure study on an A/S-type CeCu$_2$Si$_2$ single crystal \cite{lengyel11} revealed a line of QCPs in the normal state, i.e., in the zero-temperature magnetic field -- pressure plane. This line hits the superconducting phase boundary (upper critical field vs. pressure) where the latter exhibits its maximum. It can be extended into the superconducting state and, thus, demonstrates a close connection of magnetism and superconductivity.

\subsection*{Antiferromagnetic order}
Initial neutron diffraction experiments were performed on powder samples due to the unavailability of sufficiently large single crystals at that time. Incommensurate antiferromagnetic order was detected in CeCu$_2$(Si$_{1-x}$Ge$_x$)$_2$ for $x\ge 0.6$ \cite{knebel96,krimmel97,krimmel97a}. Subsequent single crystal neutron diffraction experiments on CeCu$_2$(Si$_{1-x}$Ge$_x$)$_2$ with lower Ge content down to $x = 0.05$ revealed that the magnetic structure is incommensurate for the whole doping series with a wave vector $\bm Q$ of the antiferromagnetic order (with respect to a nearby nuclear Bragg reflection) being close to $\bm Q = (0.25~0.25~0.5)$ \cite{stockert03,faulhaber04,stockert05,faulhaber06,faulhaber08}. While the $Q_L$ component of $\bm Q = (Q_H~Q_H~Q_L)$ remains almost constant for all Ge concentrations with $Q_L \approx 0.5$, $Q_H$ changes from $Q_H \approx 0.27 - 0.28$ for $x > 0.25$ to $Q_H \approx 0.22$ for lower Ge content \cite{stockert05}. Moreover, CeCu$_2$(Si$_{1-x}$Ge$_x$)$_2$ with $x > 0.25$ exhibits spin-reorientation and lock-in transitions within the magnetically ordered state. At high temperatures just below the N\'eel temperature the propagation vector in CeCu$_2$(Si$_{1-x}$Ge$_x$)$_2$ is slightly temperature dependent. However, at the lock-in transition, the magnetic structure changes through a first-order transition with a jump of the propagation vector to a fixed spin structure with no further change of the propagation vector towards lower temperatures. Here, the low-temperature phase in CeCu$_2$(Si$_{0.55}$Ge$_{0.45}$)$_2$ has been studied in detail with a combination of magnetic x-ray scattering and high-resolution elastic neutron scattering \cite{faulhaber08}. The measurements revealed that the magnetic structure breaks the body-centering of the crystallographic structure below the lock-in transition and is best described by two counterrotating cycloids \cite{faulhaber08}. At ambient pressure the ordered moment $\mu_{ord} \approx 0.5\,\mu_{\rm B}$ for CeCu$_2$(Si$_{0.55}$Ge$_{0.45}$)$_2$ is already substantially reduced compared to that of CeCu$_2$Ge$_2$.  In general, the ordered moment and the N\'eel temperature in CeCu$_2$(Si$_{1-x}$Ge$_x$)$_2$ decrease for lower Ge concentration towards CeCu$_2$Si$_2$.

\begin{figure}[tb]
\centering
\includegraphics[width=\linewidth,clip]{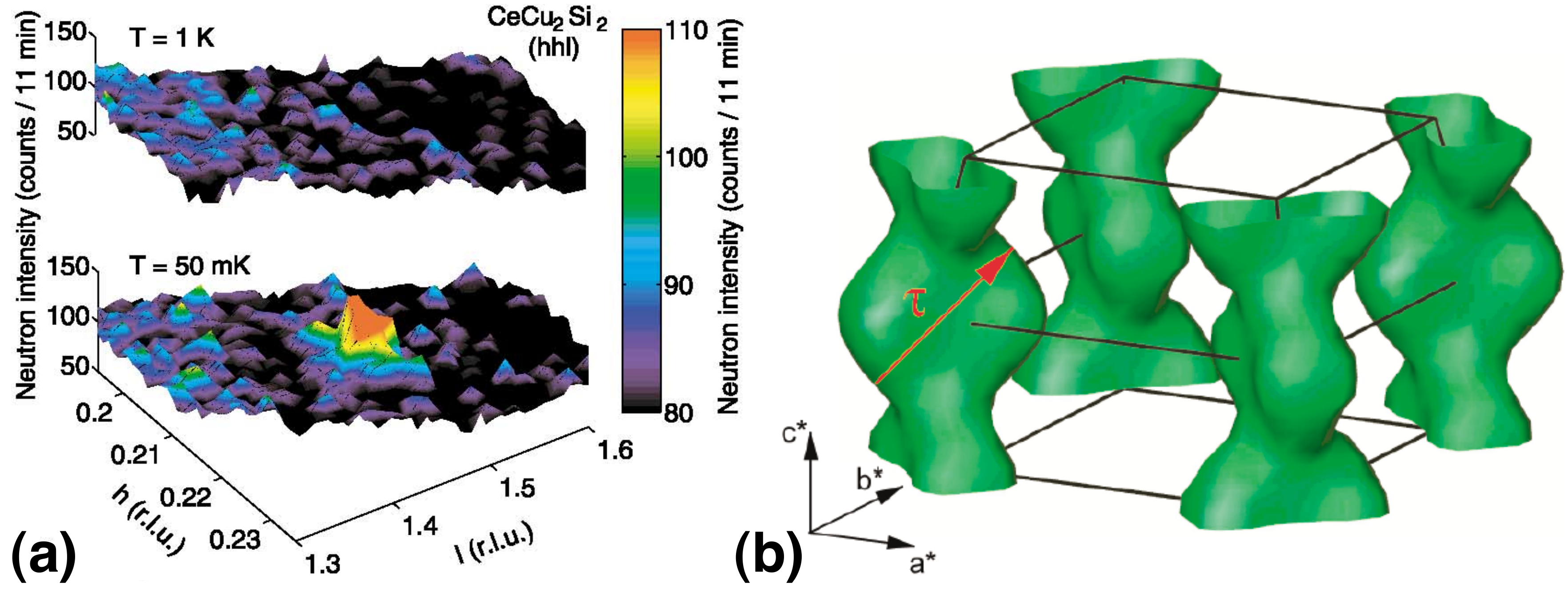}
\caption{(Color online) (a)  Intensity map of the reciprocal $(h h l)$ plane around $\bm q = (0.21~0.21~1.45)$ CeCu$_2$Si$_2$ at $T = 50$\,mK and $1$\,K \cite{stockert04}. (b) Corrugated Fermi surface of the heavy quasiparticles in CeCu$_2$Si$_2$ as derived from calculations using the renormalized band method. The vector $\tau$ denotes a nesting wave vector of the renormalized Fermi surface. \cite{stockert04}.\label{fig5}}
\end{figure}

The nature of the antiferromagnetic order in pure CeCu$_2$Si$_2$ was determined to be an incommensurate spin-density wave \cite{stockert04,stockert05,sparn06,stockert06a,stockert07a}. Initial neutron diffraction experiments were performed on A-type CeCu$_2$Si$_2$ \cite{stockert04,stockert06a} and later confirmed on an A/S-type single crystal \cite{stockert07a}. As displayed in Fig.\,\ref{fig5}(a) intensity maps of reciprocal space were recorded at different temperatures in order to search for magnetic superstructure peaks. Antiferromagnetic peaks are clearly visible at temperatures below $T_{\rm N} \approx 0.8$\,K but disappear above $T_{\rm N}$. Their incommensurate positions are described by a propagation vector $\bm Q = (0.215~0.215~0.530)$ at $T = 0.05$\,K \cite{stockert04}. While at high temperature the propagation vector shows temperature dependence, the magnetic structure locks in to a structure with constant propagation vector below $T = 0.3 - 0.35$\,K via a first order transition. The ordered magnetic moment is estimated to be $\mu_{ord} \approx 0.1\,\mu_{\rm B}$ assuming a similar structure as for the Ge substituted samples.
Band-structure calculations using a renormalized band method yielded the Fermi surface of the heavy quasiparticles shown in Fig.\,\ref{fig5}(b) \cite{stockert04}. The strong corrugation gives rise to nesting with a wave vector $\tau$ being almost identical to the observed propagation vector $\bm Q$ of the magnetic order. In addition, the wave-vector dependent susceptibility, calculated for non-interacting quasiparticles, shows a maximum at the wave vector $\bm q = \bm Q$. 
When the Coulomb repulsion between the f-electrons is large but finite, we expect a strongly reduced but nonzero
onsite component of the effective exchange interaction.
When this $U_{\rm eff}$ is combined with $I_{\bf q}$, given in Eq.~(\ref{QS:eq:KLHamiltonian}) (see below, Sec.\,6), there
will be an RPA-type enhancement of the susceptibility of $\chi_0(\bm q,\omega)$ to $\chi(\bm q,\omega)$ at $\bm q = \bm Q$.
The nesting wave vector $\bm Q$ then specifies the ordering wave vector.

\subsection*{Competition between long-range antiferromagnetism and superconductivity}
Several experiments were devoted to studying the interplay between superconductivity and antiferromagnetic order. Neutron scattering measurements were performed on an A/S-type CeCu$_2$Si$_2$ single crystal which shows a transition into the antiferromagnetically ordered state at $T_{\rm N} \approx 0.7$\,K and a superconducting phase below $T_c \approx 0.55$\,K. Inside the antiferromagnetic phase magnetic order with the same propagation vector has been detected as for A-type samples, but the magnetic intensity has completely vanished in the superconducting phase \cite{stockert07a,arndt10}. Hence, it can be concluded that long-range antiferromagnetism and superconductivity in CeCu$_2$Si$_2$ exclude each other on a microscopic scale. Neutron scattering results together with $\mu$SR experiments provide evidence that the superconducting volume fraction increases on the expense of the antiferromagnetic volume when lowering the temperature through $T_c$, without coexistence of both phenomena in the same volume \cite{stockert07a,stockert06b}. As mentioned before, the superconducting transition at $T_c$ in A/S-type CeCu$_2$Si$_2$ has been confirmed to be first order \cite{stockert06b}.

Another set of neutron scattering experiments focussed on an A/S-type CeCu$_2$Si$_2$ single crystal in which the antiferromagnetic and the superconducting transitions are almost degenerate, i.e., $T_{\rm N} \approx T_c \approx 0.7$\,K \cite{faulhaber07}. The neutron measurements again clearly indicate that also in this crystal  magnetic order and superconductivity compete and exclude each other in the same volume. Instead, phase separation into superconducting and antiferromagnetic volumes is present.

The neutron experiments on A/S-type CeCu$_2$Si$_2$ indicate that the same 4f electrons of cerium are involved in the antiferromagnetic order as well as in the unconventional superconducting state. The latter fact was already concluded from the first heat capacity measurements on CeCu$_2$Si$_2$ which revealed a jump height in $C/T$ at $T_c$ being comparable to the huge normal state value of $C/T$ at $T_c$ \cite{steglich79}.


\subsection*{Superconducting state}
The magnetic response in S-type CeCu$_2$Si$_2$ was studied by high-resolution inelastic neutron scattering in the normal as well as the superconducting state \cite{stockert08,stockert11,arndt11}. In detail the momentum and energy dependence of the spin excitations were measured throughout the relevant part of the Brillouin zone around the antiferromagnetic wave vector $\bm Q$. The aim of the measurements was twofold: first, to characterize the magnetic excitation spectrum and to calculate the magnetic exchange energies and compare the results to the condensation energy, and second, to verify the vicinity of this compound to a quantum critical point. S-type CeCu$_2$Si$_2$ becomes superconducting below $T_c \approx 0.6$\,K and shows an upper critical magnetic field $B_{c2} \approx 1.7$\,T necessary to suppress superconductivity. No long-range antiferromagnetic order has been detected in S-type CeCu$_2$Si$_2$. 

\begin{figure}[t]
\centering
\includegraphics[width=\linewidth,clip]{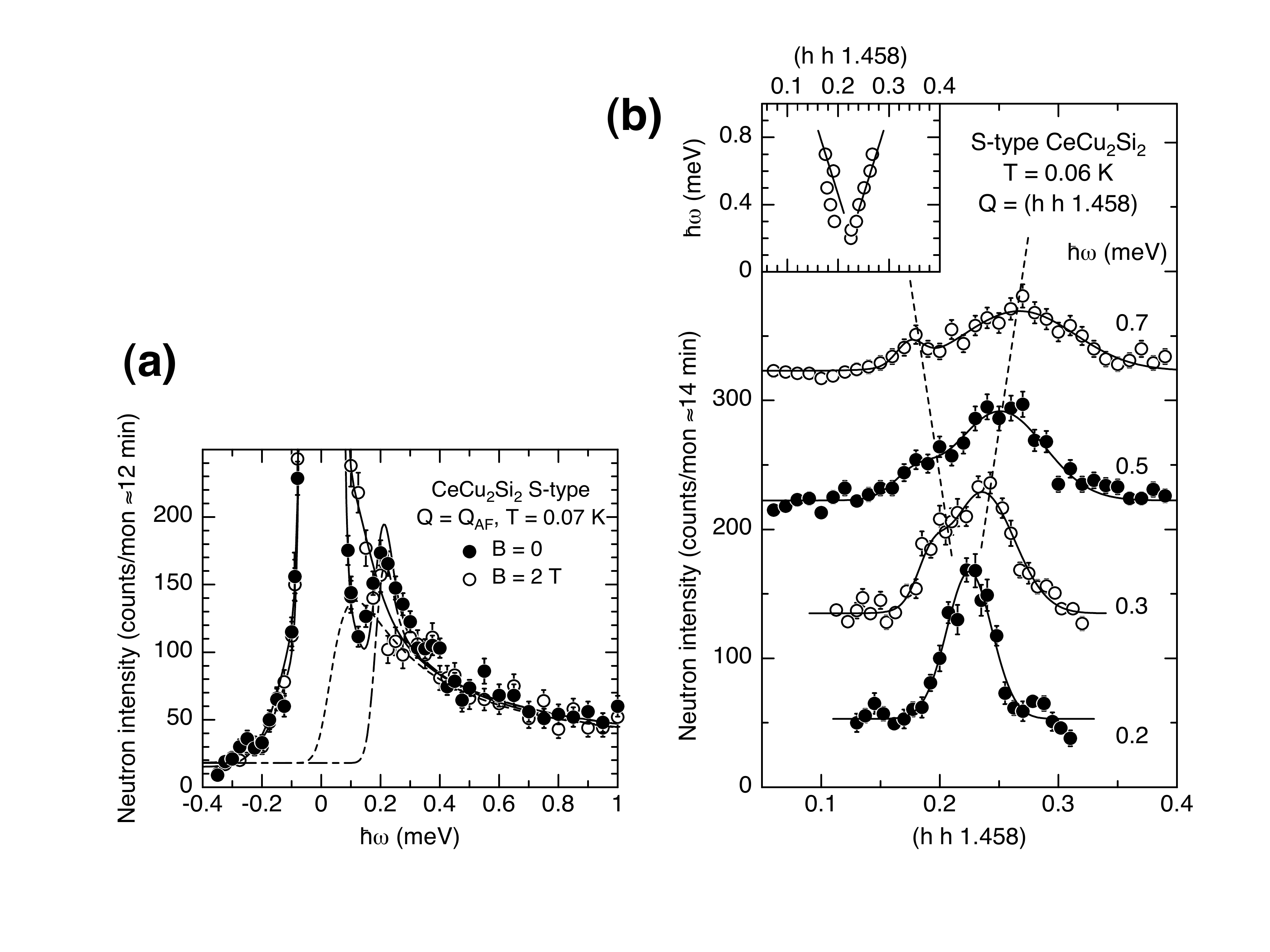}
\caption{(a) Energy scans in S-type CeCu$_2$Si$_2$ at $\bm Q = (0.215~0.215~1.458)$ in the superconducting state at $T = 0.07$\,K, $B = 0$ and in the normal state at $T = 0.07$\,K, $B = 2$\,T  \cite{stockert11}. (b) Wave vector dependence of the magnetic response in S-type CeCu$_2$Si$_2$ around the antiferromagnetic $\bm Q$ at $T = 0.06$\,K for several energy transfers $\hbar\omega$. Inset: Dispersion of the magnetic response as deduced from the scans displayed in the main panel \cite{stockert11}.\label{fig6}}
\end{figure}

Energy scans have been recorded at the antiferromagnetic wave vector $\bm Q$ (in the vicinity of (002)) in the normal and superconducting states of S-type CeCu$_2$Si$_2$, as displayed in Fig.\,\ref{fig6}(a)~\cite{stockert11}. Here, measurements in the superconducting state were performed in zero magnetic field at $T = 0.07$\,K, i.e., well below $T_c$, while the normal state has been generated by applying an overcritical magnetic field ($B = 2$\,T) at lowest temperatures. The response consists of a strong, incoherent elastic signal and the magnetic response. The magnetic fluctuations in the normal state are quasielastic and can be well described by a simple Lorentzian line multiplied by the Bose factor (dashed line). It should be noted that irrespective of how the normal state was approached, i.e., either by application of a magnetic field or by increasing the temperature above $T_c$, the magnetic response turns out to be quasielastic. In contrast, the response in the superconducting state appears to be gapped, and spectral weight is transferred from low energies to energies above the spin excitation gap which attains a value of $\hbar\omega_{gap} \approx 0.2$\,meV at $T = 0.07$\,K. This corresponds to $\hbar\omega_{gap} \approx 3.9$\,k$_{\rm B}T_c$ and is roughly 10\% smaller than the value predicted for a weak-coupling d-wave superconductor \cite{ohkawa87} and about 20\% below $5$\,k$_{\rm B}T_c$ as inferred from Cu-NQR measurements \cite{ishida99,fujiwara08}. With increasing temperature the spin gap becomes smaller and closes at $T_c$.

The wave vector dependence of the magnetic response in the superconducting state is displayed in Fig.\,\ref{fig6}(b) \cite{stockert11}. The spin excitations around the antiferromagnetic wave vector $\bm Q$ are part of an overdamped dispersive mode with a mode velocity being more than an order of magnitude smaller than the renormalized Fermi velocity \cite{rauchschwalbe82}. This indicates a clear retardation in the interactions between the spin excitations and the heavy quasiparticles \cite{stockert11}. With this detailed knowledge of the energy and momentum dependence of the spin excitations in S-type CeCu$_2$Si$_2$, the difference in magnetic exchange energy between the superconducting and the normal state could be calculated \cite{stockert11}. 
The gain in exchange energy as described in section \ref{sec:theo} is larger roughly by a factor of 20 than the condensation energy \cite{stockert11}. This implies that the spin excitations are the driving force for superconductivity in CeCu$_2$Si$_2$. It also implies that there is  a large loss of kinetic energy, that can be interpreted as the result of Mott-like physics associated with the Kondo effect.

To be able to study the spin fluctuations in the normal state of CeCu$_2$Si$_2$ when tuning the system towards the quantum critical point, superconductivity was suppressed in a magnetic field $B = 1.7$\,T ($\approx B_{c2}$). At higher temperature the magnetic response at the ordering wave vector $\bm Q$ remains quasielastic, but broadens and decreases in intensity \cite{arndt11}. Conversely, the magnetic response displays a pronounced slowing down for $T \rightarrow 0$. The temperature dependence of the spin fluctuations and their scaling behavior are in line with the expectations for a three-dimensional spin-density-wave quantum critical point.

\subsection*{Valence instability}
In close analogy to the Berk-Schrieffer paramagnon scenario of magnetically mediated superconductivity, it has been argued that density
fluctuations may induce Cooper pairing~\cite{Kohn.65,Onishi.00,Monthoux.04}. Near, e.g., charge-density-wave (CDW) instabilities, density-density fluctuations are expected to be large and the associated charge susceptibility will be enhanced~\cite{Monthoux.04}. Near an abrupt change of the unit-cell volume without changing the crystal symmetry 
(``Kondo-volume collapse''), which is typically of first order, the associated charge 
susceptibility remains finite but increases with closer proximity to the critical endpoint. 
A first order valence instability is known to occur in cerium-based heavy fermions as well as in elemental cerium
where the 4f electron delocalized towards the high-pressure side. 
The first direct indications for a valence transition in CeCu$_2$Si$_2$ under pressure come from  $L_{III}$ x-ray absorption spectroscopy~\cite{roehler88,rueff11}.
In elemental cerium, the critical endpoint occurs
at very high temperatures. In  contrast, in  the series CeCu$_2$(Si$_{1-x}$Ge$_{x}$)$_2$ 
($x=0.01,0.05,0.1$) and CeCu$_2$Si$_2$ the
critical endpoint is shifted down to $10-20$\,K, opening the way for sufficiently large  density-density fluctuations to drive superconductivity~\cite{yuan06}. 
The hallmark of this quantum phase transition in CeCu$_2$(Si$_{1-x}$Ge$_{x}$)$_2$  at a pressure of around $4$\,GPa is the linear-in-temperature resistivity that appears to be disorder independent and the colossal pressure dependence of the residual resistivity close to the transition~\cite{yuan06}.
Heat capacity and electrical resistivity under pressure in CeCu$_2$Si$_2$ point to  $p_c^*\approx 4.5$\,GPa as the pressure where the first-order quantum phase transition occurs~\cite{Holmes.07}.
Near this pressure, the ground state of the  series CeCu$_2$(Si$_{1-x}$Ge$_{x}$)$_2$  ($x=0,0.01,0.05,0.1$) is superconducting and at $p_c$ an increased jump of the specific heat has been observed~\cite{Holmes.07}.
X-ray diffraction under pressure and at $10$\,K has established that in CeCu$_2$Ge$_2$ an anomalous volume contraction
occurs at a pressure around $15$\,GPa which is close to the pressure where $T_c$ is maximal ($\approx 16$\,GPa)~\cite{onodera02}.
The enormous enhancement of the residual resistivity around the pressure of the valence transition together with
the linear-in-temperature behavior of the resistivity are strong indications that the superconductivity in proximity to $p_c^*$ is driven by nearly critical valence fluctuations \cite{Holmes.07}.
Miyake and co-workers have considered a microscopic model that includes the Coulomb interaction between the 4f-electron and the itinerant electrons~\cite{Onishi.00}. 
Based on a slave boson calculation in the large-N limit (N being the spin degeneracy) and Gaussian fluctuations around the saddle point, it was shown that a superconducting pairing interaction ensues which is almost constant in momentum space. This is in line with expectations for a valence transition which leaves the crystal structure invariant.
For a recent review on the effects of valence fluctuations in Ce- and Yb-based heavy fermion systems, see~\cite{Watanabe.11}.

As a result, the series CeCu$_2$(Si$_{1-x}$Ge$_{x}$)$_2$  ($x=0,\ldots,1$) is characterized by two superconducting phases of different origin (cf. Fig.\,\ref{fig4}). Near the antiferromagnetic quantum critical point, superconductivity is magnetically driven whereas the superconducting
phase near the valence transition seems to be mediated by local density fluctuations.
The two superconducting domes are merged in pure CeCu$_2$Si$_2$ and CeCu$_2$Ge$_2$ yielding one extended pressure range where superconductivity is found, they have been separated into two superconducting regions in Ge-doped CeCu$_2$(Si$_{1-x}$Ge$_x$)$_2$~\cite{yuan03}.

\subsection*{Order parameter symmetry}
For electronically driven
superconductors, a standard s-wave order parameter is not favored since the  pairing potential shows strong momentum dependence across the Fermi surface.
All 122 superconductors discovered so far are even-parity superconductors and the most likely order parameter symmetry is d-wave. CeCu$_2$Si$_2$ is so far the only 122 system where the order parameter symmetry has been investigated in 
detail.  Heat capacity measurements performed above $T_c/2$ suggested that the superconducting pairing symmetry changes with pressure from $d_{x^2-y^2}$ in A/S-type CeCu$_2$Si$_2$ to $d_{xy}$ in the valence driven superconducting phase \cite{lengyel09}.
Heat-capacity measurements on S-type CeCu$_2$Si$_2$ were inconclusive \cite{stockert11}. At temperatures right below T$_c$, they also point to a $d_{x^2-y^2}$ order parameter. As one lowers the temperature further, the data are more compatible with $d_{xy}$  as the
pairing symmetry.
In a recent experiment on S-type CeCu$_2$Si$_2$, angle resolved $H_{c2}$ measurements were performed which clearly point to a $d_{xy}$ symmetry of the superconducting order parameter~\cite{vieyra11}.
The spin fluctuation spectrum of superconducting S-type CeCu$_2$Si$_2$ near the critical pressure has been investigated in detail~\cite{stockert11}. The opening of the superconducting gap is accompanied by an enhancement
of the spin fluctuations, see Fig.~\ref{fig6}(a). 
This is unlike the sharp spin resonance seen in, e.g., CeCoIn$_5$ or the cuprates~\cite{stock08}.
The spin resonance was taken as an indication for a $d_{x^2-y^2}$ order parameter~\cite{eremin08}.
For CeCoIn$_5$, however, a different interpretation of the spin resonance has been made by considering the
strongly three-dimensional character of the Fermi surface\cite{Chubukov.08}.

\section{Superconductivity, magnetism and hidden order in URu$_2$Si$_2$}

One of the early discovered heavy-fermion superconductors is URu$_2$Si$_2$ \cite{schlabitz84,schlabitz86}, 
with $T_c \approx 1.5$\,K - substantially higher than $T_c$ of its then known Ce- and U-based counterparts CeCu$_2$Si$_2$ ($T_c = 0.6$\,K), UBe$_{13}$ ($T_c \approx 0.9$\,K) and UPt$_3$ ($T_c \approx 0.5$\,K). In URu$_2$Si$_2$, superconductivity forms well below $T_0 = 17.5$\,K, signaling a second-order phase transition, with a large mean-field type phase transition anomaly in the specific heat \cite{schlabitz86,palstra85,maple86}. From the exponential temperature dependence of the specific heat at $T < T_0$, the opening of a gap $\Delta \approx 0.1$\,meV in the single-electron excitations was inferred. In addition, very different values for the Sommerfeld coefficient of the electronic specific heat have been estimated (as $T \rightarrow 0$) above and below $T_0$, i.e., $\gamma \approx 180$ and $70$\,mJ/K$^2$mol, respectively. This led to the assumption of (i) a removal of approximately $60$\% of the renormalized Fermi surface \cite{maple86} and, taking into account also the "Cr-type" shape of $\gamma(T)$ near $T_0$ \cite{schlabitz84,maple86}, (ii) the formation of antiferromagnetic (spin-density-wave) order at $T \le T_0$. However, in proportion to the very large removal of entropy associated with the continuous phase transition at $T_0$, the ordered moment found by neutron diffraction was much too small: $\mu{ord} \approx 0.03\,\mu_{\rm B}$/U \cite{broholm87}. A number of theoretical proposals concerning the nature of the ordered low-temperature phase of URu$_2$Si$_2$ appeared during the last two decades, e.g., singlet magnetism \cite{barzykin95}, multipole \cite{santini94,ohkawa99,kuramoto06} or orbital ordering \cite{chandra02}, charge-density wave coupled to some antiferromagnetic order \cite{mineev05} as well as helical spin order \cite{varma06}. It is fair to say that the nature of this phase is still unknown today, and it is commonly called "hidden order". Of course, this is a misnomer, as this phase is not hidden but of a nature that could  not yet be resolved. 

Inelastic neutron scattering results \cite{broholm87,wiebe07} indicate the existence of gapless overdamped in-plane magnetic excitations at the incommensurate wave vector $\bm Q^* = (1\pm0.4,0,0)$ which become gapped ($\hbar\omega_{gap} \approx 4$\,meV) and remain well defined (up to $\approx 10$\,meV) below $T_0$. Similar spin dynamics has been observed earlier for the prototypical spin-density-wave metal Cr \cite{fawcett88}. Angle-resolved photoemission measurements \cite{santander-syro09} on the other hand, revealed a light band which crosses $E_{\rm F}$ near $(0,0,\pm0.3)$ as well as $(\pm0.3,0,0)$ above $T_0 = 17.5$\,K, and transforms into a heavy band below $T_0$. In addition, magnetic excitations at a commensurate wave vector $\bm Q_0 = (1,0,0)$ were observed which form a gap of $\approx 1.6$\,meV below $T_0$ \cite{broholm87,wiebe07}. 

High-resolution scanning tunneling microscopy/spectroscopy studies \cite{schmidt10,aynajian10} revealed asymmetric Fano-type resonances, which typically arise in Kondo systems. They show the same periodicity as the U lattice and develop a gap below $T_0$. This gap feature is described by a mean-field-type order parameter of the hidden order  phase in Ref. \cite{aynajian10}. In Ref. \cite{schmidt10}, on the other hand, it is related, via imaging heavy-quasiparticle interference, to a rapid splitting of a light band into two new heavy-fermion bands, as was predicted \cite{martin82,kim90}. 
Recent magnetic torque measurements under rotation of the magnetic field within the basal plane revealed an in-plane
anisotropy of the magnetic susceptibility below $T_0$~\cite{okazaki11}. The four-fold rotational symmetry of
the tetragonal ThCr$_2$Si$_2$ structure is spontaneously broken at $T_0$ as reflected by two-fold oscillations
of the magnetic torque under in-plane rotation of the field. These observations suggest that the hidden order  phase is an
electronic ``nematic'' phase, i.e., a metallic phase with translational symmetry preserved but spontaneously broken
rotational symmetry, as discovered earlier for, e.g., Sr$_3$Ru$_2$O$_7$~\cite{Borzi.07}.
The hidden-order gap is the subject of a recent first-principle theoretical treatment \cite{haule09}, in which a $5f^2$ (U$^{4+}$) configuration with CEF splitting is assumed. Here, the hidden order  was described by a multichannel Kondo effect which becomes arrested upon cooling to below $T_0 = 17.5$\,K. However, at present, no experimental evidence for CEF splitting has been reported for this compound with $T_{\rm K} \approx 60$\,K \cite{kohara86}. More generally, the action of the Kondo effect in U-based heavy fermions is conceptually debated \cite{cox98,fujimori11}, 
mainly because of the more delocalized character of the $5f$ electrons compared to the $4f$ 
electrons. In short, the hidden order  phase remains enigmatic. 

\begin{figure}[t]
\centering
\includegraphics[width=0.85\linewidth,clip]{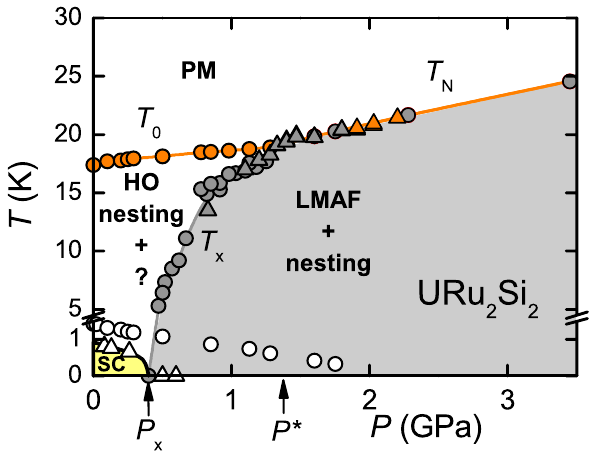}
\caption{(Color online) Temperature - pressure phase diagram of URu$_2$Si$_2$ as deduced from resistivity (circles) and heat capacity (triangles) measurements (HO: hidden order, LMAF: large moment antiferromagnetic order, PM: paramagnetic phase, SC: superconductivity) \cite{hassinger08}.\label{fig9}}
\end{figure}

As displayed in Fig.\,\ref{fig9} at finite pressure, hidden order  becomes replaced by strong antiferromagnetic order ($\mu_{ord} \approx 0.4\,\mu_{\rm B}$) with wave vector $\bm Q_0 = (1,0,0)$ \cite{amitsuka07}, and the boundary between these two phases was found to stretch from $p_c \approx 0.5$\,GPa (as $T \rightarrow 0$) to a (possible) tricritical point ($\approx 1.2$\,GPa, $\approx 18$\,K) \cite{amitsuka07}. The "weak" antiferromagnetic order, which seems to occur within the hidden order  phase, is most likely due to static antiferromagnetically ordered regions that exist in  a small part of the sample, phase separated from the surrounding hidden order  region \cite{hassinger08}. Thus, d-wave superconductivity~\cite{Kasahara.07,Yano.08} seems to microscopically coexist with the hidden order  but not with antiferromagnetism \cite{hassinger08}.

\section{Theoretical considerations}

\label{sec:theo}
The interplay between quantum criticality and superconductivity
is of interest to a variety of strongly correlated systems. For antiferromagnetic heavy fermion metals,  studies in recent years on the Kondo lattice models have revealed two types of quantum critical points. 
One is a spin-density-wave  quantum critical point~\cite{hertz76,Moriya,millis93}, 
where the Kondo effect remains intact  and the critical modes are 
fluctuations of the antiferromagnetic order parameter. The other one is the local quantum critical point~\cite{si01,coleman01}, where the Kondo effect breaks down. These are described in terms of a Kondo-breakdown energy scale,
$E_{\rm loc}^*$, which remains nonzero at a spin-density-wave quantum critical point but vanishes at the local quantum critical point.

\begin{figure}[t]
\centering
\includegraphics[width=0.85\linewidth,clip]{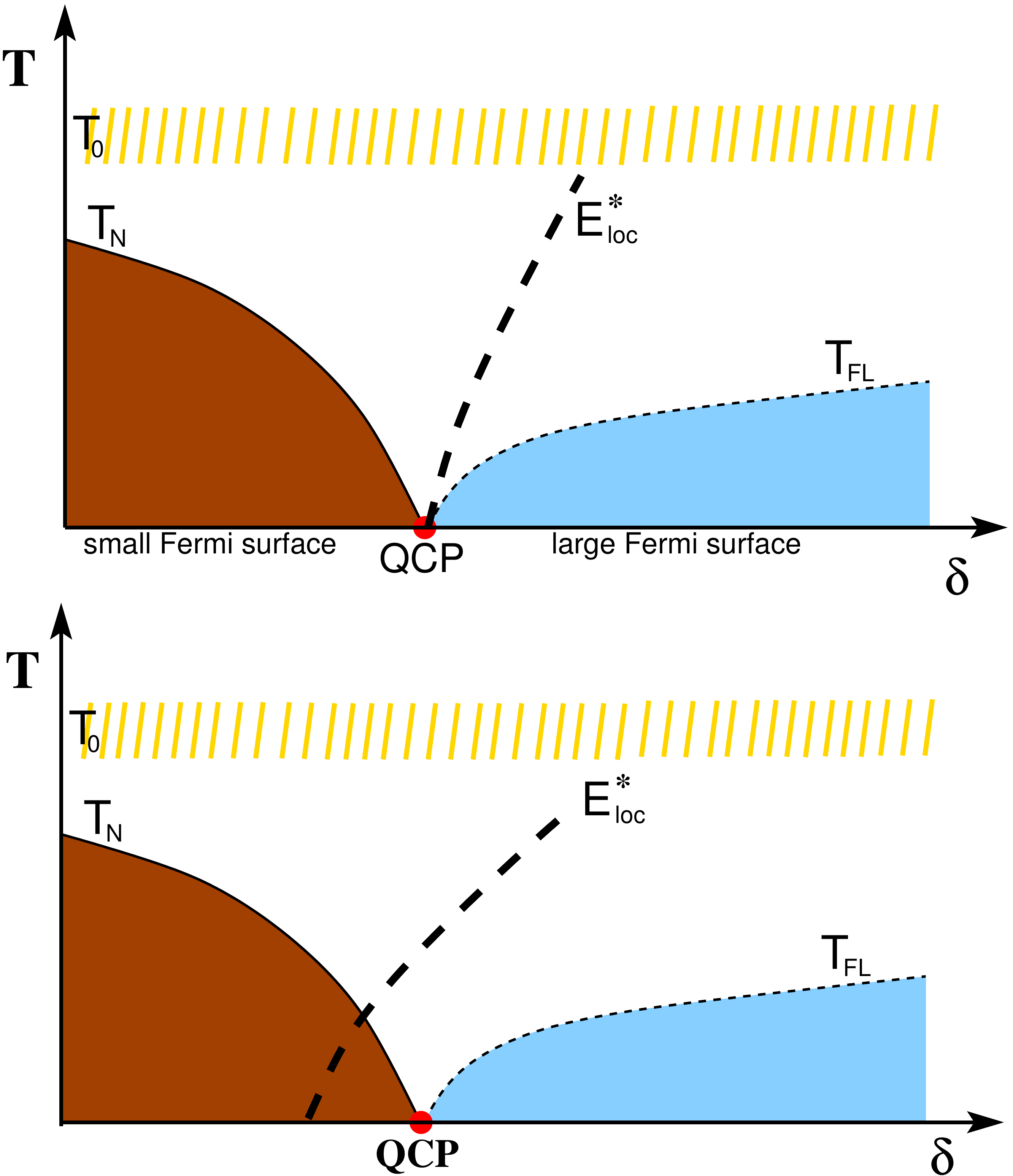}
\caption{(Color online) Two types of quantum critical points (QCPs). Schematic temperature -- control parameter $T-\delta$ phase diagrams for local criticality (upper panel) and spin-density-wave criticality (lower panel). For details see main text.
\label{fig_theory1}}
\end{figure}

Fig.~\ref{fig_theory1} illustrates the phase diagrams in the temperature-control parameter space. The control parameter is specified in terms of the ratio of the Kondo temperature to the RKKY interaction, defined for a Kondo lattice Hamiltonian,
\begin{eqnarray}\label{QS:eq:KLHamiltonian}
H_{\rm KL}=
 \sum_{ ij } t_{ij} \sum_{\sigma}
c^{\dagger}_{i\sigma}
c^{\phantom\dagger}_{j\sigma}
+
 \sum_{ ij } I_{ij}
{\bm S}_i \cdot {\bm S}_j
+  \frac{1}{2} \sum_{i} J_K {\bm S}_i \cdot c^{\dagger}_{i}
{\bm \sigma} c^{\phantom\dagger}_{i} .
\label{QS:kondo-lattice-model}
\end{eqnarray}
It contains a conduction-electron band, $c_{i\sigma}$,
whose hopping matrix $t_{ij}$ specifies 
a band dispersion $\varepsilon^{\phantom\dagger}_{\bf k}$.
The spins of the conduction
electrons, ${\bm s}_{c,i} = (1/2) c_{i}^{\dagger} {\bm\sigma} c_i$,
are coupled to the spin of the local moment, $\bm {S}_i$,
via an antiferromagnetic
Kondo exchange interaction $J_{\rm K}$. The corresponding bare Kondo temperature is 
$
T_{\rm K}^0 \approx \rho_0^{-1} \exp(-1/\rho_0J_{\rm K})
$, where $\rho_0$ is the conduction electron density of states at the Fermi energy.
Finally, the antiferromagnetic RKKY interaction has the typical scale $I$.

When $\delta = T_{\rm K}^0/I$ is large, the Kondo interaction drives the formation of Kondo singlets between
the local moments and conduction electrons. At high temperatures,
the system is in a fully incoherent regime with the local moments
weakly coupled to conduction electrons. Going below the scale
$T_0$, the initial screening of the local moments starts to
set in. Eventually, at temperatures
below a Fermi-liquid scale, $T_{\rm FL}$, the heavy 
quasiparticles
are fully developed.

Decreasing $\delta$ enhances the tendency of antiferromagnetic correlations. 
When the RKKY 
 interaction among the local moments
becomes larger than
the Kondo interaction, the system is expected to develop
antiferromagnetic order. An antiferromagnetic quantum critical point is then to be expected
when $\delta = T_{\rm K}^0/I$
reaches some
critical value $\delta_c$. At 
$\delta < \delta_c$, 
the antiferromagnetic order
will develop as 
the 
temperature is lowered through the antiferromagnetic-ordering
line, $T_{\rm N}(\delta)$.

In addition, 
the RKKY interactions will eventually lead
to the suppression of the Kondo singlets. Qualitatively,
RKKY interactions promote singlet formation among
the local moments, thereby reducing the tendency of singlet
formation between the local moments and conduction electrons.
This will define an energy ($E_{\mathrm{loc}}^*$) 
or temperature ($T_{ }^*$)
scale, describing the breakdown of the Kondo effect. 
The $T_{ }^*$ line represents a crossover
at finite temperatures, but turns into
a sharp transition at zero temperature.
The notion of Kondo breakdown in quantum critical heavy fermions
was introduced
in the theory of local quantum
criticality~\cite{si01,coleman01} and also appeared in 
subsequent work~\cite{senthil04,paul07}.
The Kondo breakdown effect is 
alternatively referred to as 
a Mott localization of the 
$f$-electrons.

For a spin-density-wave quantum critical point in heavy fermions, quantum criticality at asymptotically low energies is described in terms of order-parameter fluctuations. These are the antiferromagnetic spin fluctuations, which have been proposed to serve as a glue for  superconductivity \cite{schmitt-rink86,scalapino86}. There have been indirect indications that $E_{\rm loc}^*$ is nonzero in CeCu$_{\rm 2}$Si$_{\rm 2}$. The low temperature/energy properties have been shown to be compatible with the three-dimensional spin-density-wave quantum criticality. These include the temperature dependences of the
specific heat and electrical resistivity \cite{gegenwart98}, as well as dynamical spin susceptibility \cite{arndt11}. The effect of quantum criticality on superconductivity was recently studied in a quantitative way. 
The lowering of the magnetic exchange energy in the superconducting state is 
determined from the measured dynamical spin susceptibility in the normal and superconducting states~\cite{scalapino98,leggett98}:
\begin{eqnarray}
\Delta E_x^{} &\!\!=\!\!&
\frac{1}{g^2 \mu_B^2}\int_0^{\infty} \frac{d (\hbar\omega)}{\pi} \big 
[ n(\hbar \omega)+1 \big ] \times  \\
 \Bigg < &\!\!\!\!\!I\!\!\!\!\!\!&\!\!\!\!\!({\bf q})\,
\Big[\mbox{Im}\chi^{S}(q_x,q_y,q_z,\hbar\omega)-\mbox{Im} \chi^{N}(q_x,q_y,q_z,\hbar\omega)\Big ]
\Bigg >,\nonumber
\label{Ex}
\end{eqnarray}
where $<>$ denotes an average over the first Brillouin zone, and $\chi^{S/N}_{}$ is the spin susceptibility
summed over its three components. 
This is found to be larger than the superconducting condensation energy,
\begin{eqnarray}
\Delta E_C = U_N (T=0)  - U _S (T=0),
\label{Ec}
\end{eqnarray}
defined as the difference in internal energy between the (putative) normal and the superconducting states at T=0. As an immediate consequence, the antiferromagnetic excitations are seen as the primary driving force for superconductivity in CeCu$_{\rm 2}$Si$_{\rm 2}$. 

\begin{figure}[t!]
\centering
\includegraphics[width=0.9\linewidth,clip]{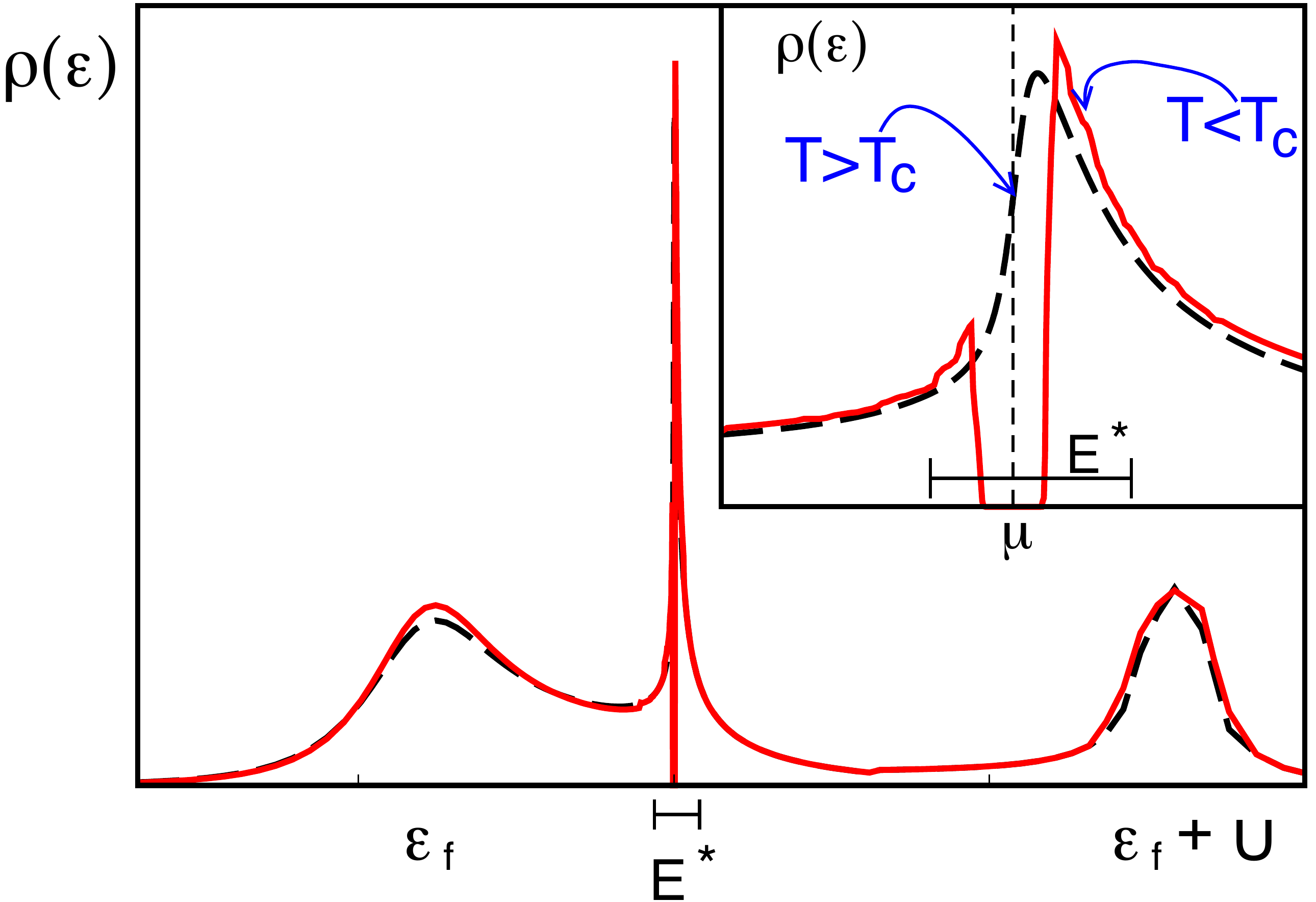}
\caption{(Color online) Schematic curves of the density of states $\rho(\epsilon)$ of the
f-electrons vs. energy in the normal (dashed line) and superconducting (solid line) states of CeCu$_2$Si$_2$ indicating the transfer of single-electron spectral weight from low energies in the normal state to energies above $E^*$ in the superconducting state.
The inset enlarges the small energy range.
\label{fig_theory2}}
\end{figure}

The analysis, however, reveals something more. It is found that $\Delta E_x $ is about 20 times of $\Delta E_C$, implying  a large kinetic energy loss,
about  19 times of $\Delta E_C$. We can discuss the origin of this effect by 
recognizing that kinetic energy in Kondo-lattice systems is primarily associated with the Kondo effect.
  As superconducting pairing in CeCu$_2$Si$_2$
occurs in the spin-singlet channel, the opening of the superconducting
gap weakens the Kondo-singlet formation and,
by extension, reduces the spectral weight of the Kondo resonance at low energies. The latter implies that, going into the superconducting state, single-electron spectral weight will be transferred from low energies to above the Kondo-breakdown scale $E_{\rm loc}^*$. This is illustrated in Fig.~\ref{fig_theory2}.

The case for superconductivity near a local quantum critical point is best made in the compound CeRhIn$_5$. Quantum-oscillation measurements show that the pressure-induced antiferromagnetic to non-magnetic quantum critical point involves a sudden reconstruction of the Fermi surface~\cite{shishido05}, from small (f-electrons localized) to large (f-electrons delocalized). At the transition, the cyclotron mass, derived from de Haas-van Alphen measurements, shows a tendency of divergence; the latter is further supported by a strong increase of the $A$-coefficient of the $T^2$ term in the resistivity as the quantum critical point is approached~\cite{knebel08}. All these provide evidence that the antiferromagnetic transition is locally quantum critical, involving a Kondo breakdown. Near this quantum critical point, superconductivity appears with a large $T_c \approx 2.2$\,K. CeRhIn$_5$ is, therefore, a prime candidate of superconductivity associated with a local quantum criticality.

\section{Conclusions and Perspectives}
We have reviewed the occurrence of heavy-fermion superconductivity in the Ce-based 122 systems and URu$_2$Si$_2$. The former materials may be divided into the local-moment, heavy-fermion antiferromagnets CeCu$_2$Ge$_2$, CeRu$_2$Si$_2$, and CePd$_2$Si$_2$ with $T_{\rm N} / T_{\rm K} \approx 1$, which are pressure-induced superconductors, and the itinerant systems CeNi$_2$Ge$_2$ and CeCu$_2$Si$_2$ with $T_{\rm N} / T_{\rm K} \ll 1$ \cite{grewe91}. The latter compounds are superconductors already at ambient pressure.

We have summarized the evidence for the spin-density-wave nature of the antiferromagnetic quantum critical point in the normal state of CeCu$_2$Si$_2$, and for the d-wave symmetry of the superconductivity near its antiferromagnetic quantum critical point. At the same time, the lowering of the magnetic exchange energy associated with the superconductivity by far exceeds the superconducting condensation energy. This is unexpected based on spin fluctuations alone, and can be naturally interpreted in terms of a transfer of single-electron spectral weight to energies above a possible Kondo-breakdown $T^*$ scale and all the way to the 4f-electron Coulomb interaction scale. It would be very instructive to search for such a $T^*$ scale in CeCu$_2$Si$_2$. Furthermore, neutron scattering measurements under pressure in other heavy fermion materials displaying a potential spin-density-wave quantum critical point under pressure, such as CePd$_2$Si$_2$, would be invaluable.

At the same time, CeRhIn$_5$ under pressure is one concrete example in which evidence exists for unconventional superconductivity in the vicinity of a Kondo-breakdown local quantum critical point. It will be highly desirable to identify further examples of unconventional superconductors in this category.

Unconventional heavy-fermion superconductivity also occurs in the vicinity of other types of quantum phase transitions. We have provided examples that involve first-order antiferromagnetic transitions, first-order valence transitions as well as the enigmatic case of URu$_2$Si$_2$. Here, d-wave superconductivity with broken time-reversal symmetry occurs near an antiferromagnetic quantum phase transition at nonzero pressure \cite{Kasahara.07,Yano.08}, raising the possibility for the relevance of antiferromagnetic order to superconductivity. However, $T_c$ appears to diminish on approach to the antiferromagnetic order, and is nonzero only in the presence of the hidden order (cf. Fig.\,\ref{fig9}). This raises the alternative possibility that the superconductivity is driven by the hidden order. This is an important issue worthy of clarifying. 

New measurement techniques continue to become available to study heavy-fermion materials. A particular example is the scanning tunneling microscopy/spectroscopy, which has been used to study the hidden-order phase in URu$_2$Si$_2$~\cite{schmidt10,aynajian10} and the emergence of the local Kondo screening and the concomitant onset of lattice coherence in YbRh$_2$Si$_2$~\cite{Ernst.11}. Extending these studies to below $T_c$ as well as to the low-temperature quantum critical regime promises to shed considerable new light on the heavy-fermion physics. Along similar lines, it will be important to develop low-temperature, high-resolution angle-resolved photoemission techniques, which should lead to much new understandings of the normal and superconducting properties of heavy-fermion materials.

Theoretical studies of superconductivity in heavy-fermion metals benefits from the existence of well-defined Hamiltonians to describe the low-energy physics of these materials. These are the Kondo-lattice Hamiltonian or, more generally, the Anderson-lattice Hamiltonian. This has been important in several regards. It has allowed systematic studies on  the heavy-fermion quantum criticality. It has also allowed a detailed analysis of the lowering of the exchange energy in the superconducting state of CeCu$_2$Si$_2$.

More broadly, heavy-fermion compounds represent  a prototype case of unconventional superconductors. There are several other families of superconductors which are also located in close proximity to magnetism. A prototypical example are the iron pnictides and chalcogenides~\cite{Kamihara.08,ishida09}. Moreover, the parent iron pnictides are
bad metals with a strong correlation-induced kinetic energy suppression~\cite{qazilbash09}, and the parent 122 iron selenides are Mott insulators~\cite{fang11}. Taken together, these properties suggest that the interplay between magnetism, electron localization, and unconventional superconductivity also plays a central role in these new high $T_c$ superconductors.

\begin{acknowledgment}
The authors thank P. Gegenwart, C. Geibel, H.S. Jeevan, M. Loewenhaupt, M. Nicklas, and G. Zwicknagl for valuable discussions. The part of the work performed at the Max Planck Institute for Chemical Physics of Solids was supported by the Deutsche Forschungsgemeinschaft under the auspices of the Research Unit 960 ``Quantum Phase Transitions''.The work at Rice University was supported by 
 the National Science Foundation under
Grant No. DMR-1006985 and the Robert A. Welch Foundation Grant No. C-1411.
 S.K., Q.S., and F.S. acknowledge support in part by the National Science Foundation under Grant No. 1066293 and the hospitality of the Aspen Center for Physics.
\end{acknowledgment}

\bibliographystyle{jpsj}
\bibliography{ce122hfs}

\begin{thebibliography}{100}

\bibitem{Tinkham}
M.~Tinkham: {\em Introduction to Superconductivity} (Krieger Pub Co, 1975).

\bibitem{hewson93}
A.~C. Hewson: {\em {The Kondo problem to Heavy Fermions}} (Cambridge University
  Press, Cambridge, 1993).

\bibitem{stewart01}
G.~R. Stewart: Rev. Mod. Phys. {\bfseries 73} (2001) 797; 
{\bfseries 78} (2006) 743.

\bibitem{loehneysen07a}
H.~v.~L\"ohneysen, A.~Rosch, M.~Vojta, and P.~W\"olfle: Rev. Mod.
  Phys. {\bfseries 79} (2007) 1015.

\bibitem{gegenwart08}
P.~Gegenwart, Q.~Si, and F.~Steglich: Nat. Phys. {\bfseries 4} (2008) 186.

\bibitem{coleman05}
P.~Coleman and A.~J. Schofield: Nature {\bfseries 433} (2005) 226.

\bibitem{Si.10}
Q.~Si and F.~Steglich: Science {\bfseries 329} (2010) 1161.

\bibitem{maple94}
M.~B. Maple, C.~L. Seaman, D.~A. Gajewski, Y.~Dalichaouch, V.~B. Barbetta,
  M.~C. Andrade, H.~A. Mook, H.~G. Lukefahr, O.~O. Bernal, and D.~E.
  MacLaughlin: J. Low Temp. Phys. {\bfseries 95} (1994) 225.

\bibitem{aronson95}
M.~C. Aronson, R.~Osborn, R.~A. Robinson, J.~W. Lynn, R.~Chau, C.~L. Seaman,
  and M.~B. Maple: Phys. Rev. Lett. {\bfseries 75} (1995) 725.

\bibitem{yuan03}
H.~Q. Yuan, F.~M. Grosche, M.~Deppe, C.~Geibel, G.~Sparn, and F.~Steglich:
  Science {\bfseries 302} (2003) 2104.

\bibitem{grosche01}
F.~M. Grosche, I.~R. Walker, S.~R. Julian, N.~D. Mathur, D.~M. Freye, M.~J.
  Steiner, and G.~G. Lonzarich: J. Phys.: Condens. Matter {\bfseries
  13} (2001) 2845.

\bibitem{severing89}
A.~Severing, E.~Holland-Moritz, B.~D. Rainford, S.~R. Culverhouse, and
  B.~Frick: Phys. Rev. B {\bfseries 39} (1989) 2557.

\bibitem{abe98}
H.~Abe, H.~Kitazawa, H.~Suzuki, G.~Kido, and T.~Matsumoto: J. Magn. Magn. Mater.
 {\bfseries 177-181} (1998) 479.

\bibitem{grier84}
B.~H. Grier, J.~M. Lawrence, V.~Murgai, and R.~D. Parks: Phys. Rev. B
  {\bfseries 29} (1984) 2664.

\bibitem{severing89a}
A.~Severing, E.~Holland-Moritz, and B.~Frick: Phys. Rev. B {\bfseries 39}
  (1989) 4164.

\bibitem{loidl92}
A.~Loidl, A.~Krimmel, K.~Knorr, G.~Sparn, M.~Lang, C.~Geibel, S.~Horn,
  A.~Grauel, F.~Steglich, B.~Welslau, N.~Grewe, H.~Nakotte, F.~R. de~Boer, and
  A.~P. Murani: Annalen der Physik {\bfseries 504} (1992) 78.

\bibitem{grosche96}
F.~M. Grosche, S.~R. Julian, N.~D. Mathur, and G.~G. Lonzarich: Physica B
 {\bfseries 223-224} (1996) 50.

\bibitem{mathur98}
N.~D. Mathur, F.~M. Grosche, S.~R. Julian, I.~R. Walker, D.~M. Freye, R.~K.~W.
  Haselwimmer, and G.~G. Lonzarich: Nature {\bfseries 394} (1998) 39.

\bibitem{steeman88}
R.~A. Steeman, E.~Frikkee, R.~B. Helmholdt, A.~A. Menovsky, J.~van~den Berg,
  G.~J. Nieuwenhuys, and J.~A. Mydosh: Solid State Commun. {\bfseries
  66} (1988) 103.

\bibitem{kernavanois05}
N.~Kernavanois, S.~Raymond, E.~Ressouche, B.~Grenier, J.~Flouquet, and
  P.~Lejay: Phys. Rev. B {\bfseries 71} (2005) 064404.

\bibitem{knopp88}
G.~Knopp, A.~Loidl, R.~Caspary, U.~Gottwick, C.~D. Bredl, H.~Spille,
  F.~Steglich, and A.~P. Murani: J. Magn. Magn. Mater.
  {\bfseries 74} (1988) 341.

\bibitem{kuechler07}
R.~K\"uchler, P.~Gegenwart, C.~Geibel, and F.~Steglich: Science and Technology
  of Advanced Materials {\bfseries 8} (2007) 428.

\bibitem{fukuhara98}
T.~Fukuhara, S.~Akamaru, T.~Kuwai, J.~Sakurai, and K.~Maezawa: J. Phys. Soc. Jpn.
 {\bfseries 67} (1998) 2084.

\bibitem{knebel99}
G.~Knebel, M.~Brando, J.~Hemberger, M.~Nicklas, W.~Trinkl, and A.~Loidl: Phys.
  Rev. B {\bfseries 59} (1999) 12390.

\bibitem{gegenwart99}
P.~Gegenwart, F.~Kromer, M.~Lang, G.~Sparn, C.~Geibel, and F.~Steglich: Phys.
  Rev. Lett. {\bfseries 82} (1999) 1293.

\bibitem{grosche00}
F.~M. Grosche, P.~Agarwal, S.~R. Julian, N.~J. Wilson, R.~K.~W. Haselwimmer,
  S.~J.~S. Lister, N.~D. Mathur, F.~V. Carter, S.~S. Saxena, and G.~G.
  Lonzarich: J. Phys.: Condens. Matter {\bfseries 12} (2000) L533.

\bibitem{rosch99}
A.~Rosch: Phys. Rev. Lett. {\bfseries 82} (1999) 4280.

\bibitem{wang11}
C.~H. Wang, J.~M. Lawrence, A.~D. Christianson, S.~Chang, K.~Gofryk, E.~D.
  Bauer, F.~Ronning, J.~D. Thompson, K.~J. McClellan, J.~A. Rodriguez-Rivera,
  and J.~W. Lynn: J. Phys.: Conf. Series {\bfseries 273} (2011)
  012018.

\bibitem{steglich97}
F.~Steglich, P.~Gegenwart, R.~Helfrich, C.~Langhammer, P.~Hellmann,
  L.~Donnevert, C.~Geibel, M.~Lang, G.~Sparn, W.~Assmus, G.~Stewart, and
  A.~Ochiai: Z. Phys. B {\bfseries 103} (1997) 235.

\bibitem{braithwaite00}
D.~Braithwaite, T.~Fukuhara, A.~Demuer, I.~Sheikin, S.~Kambe, J.-P. Brison,
  K.~Maezawa, T.~Naka, and J.~Flouquet: J. Phys.: Condens. Matter
  {\bfseries 12} (2000) 1339.

\bibitem{fak00}
B.~F\aa{}k, J.~Flouquet, G.~Lapertot, T.~Fukuhara, and H.~Kadowaki: J. Phys.:
  Condens. Matter {\bfseries 12} (2000) 5423.

\bibitem{kadowaki03}
H.~Kadowaki, B.~F\aa{}k, T.~Fukuhara, K.~Maezawa, K.~Nakajima, M.~A. Adams,
  S.~Raymond, and J.~Flouquet: Phys. Rev. B {\bfseries 68} (2003) 140402.

\bibitem{godart83}
C.~Godart, L.~Gupta, and M.~Ravet-Krill: J. Less Common Metals
  {\bfseries 94} (1983) 187.

\bibitem{settai97}
R.~Settai, A.~Misawa, S.~Araki, M.~Kosaki, K.~Sugiyama, T.~Takeuchi, K.~Kindo,
  Y.~Haga, E.~Yamamoto, and Y.~\={O}nuki: J. Phys. Soc. Jpn.
   {\bfseries 66} (1997) 2260.

\bibitem{kawasaki98}
Y.~Kawasaki, K.~Ishida, Y.~Kitaoka, and K.~Asayama: Phys. Rev. B {\bfseries 58}
  (1998) 8634.

\bibitem{kawarazaki00}
S.~Kawarazaki, M.~Sato, Y.~Miyako, N.~Chigusa, K.~Watanabe, N.~Metoki,
  Y.~Koike, and M.~Nishi: Phys. Rev. B {\bfseries 61} (2000) 4167.

\bibitem{graf97}
T.~Graf, J.~D. Thompson, M.~F. Hundley, R.~Movshovich, Z.~Fisk, D.~Mandrus,
  R.~A. Fisher, and N.~E. Phillips: Phys. Rev. Lett. {\bfseries 78} (1997)
  3769.

\bibitem{movshovich96}
R.~Movshovich, T.~Graf, D.~Mandrus, J.~D. Thompson, J.~L. Smith, and Z.~Fisk:
  Phys. Rev. B {\bfseries 53} (1996) 8241.

\bibitem{grosche97}
F.~Grosche, S.~Julian, N.~Mathur, F.~Carter, and G.~Lonzarich: Physica B
 {\bfseries 237-238} (1997) 197.

\bibitem{araki02}
S.~Araki, M.~Nakashima, R.~Settai, T.~C. Kobayashi, and Y.~Onuki: J. Phys.: 
Condens. Matter {\bfseries 14} (2002) L377.

\bibitem{villaume08}
A.~Villaume, D.~Aoki, Y.~Haga, G.~Knebel, R.~Boursier, and J.~Flouquet: J. Phys.: 
Condens. Matter {\bfseries 20} (2008) 015203.

\bibitem{boursier08}
R.~Boursier, A.~Villaume, G.~Lapertot, D.~Aoki, G.~Knebel, and J.~Flouquet:
  Physica B {\bfseries 403} (2008) 726 .

\bibitem{araki01}
S.~Araki, R.~Settai, T.~C. Kobayashi, H.~Harima, and Y.~\ifmmode~\bar{O}\else
  \={O}\fi{}nuki: Phys. Rev. B {\bfseries 64} (2001) 224417.

\bibitem{araki02a}
S.~Araki, R.~Settai, M.~Nakashima, H.~Shishido, S.~Ikeda, H.~Nakawaki, Y.~Haga,
  N.~Tateiwa, T.~C. Kobayashi, H.~Harima, H.~Yamagami, Y.~Aoki, T.~Namiki,
  H.~Sato, and Y.~Onuki: J. Phys. Chem. Solids {\bfseries
  63} (2002) 1133 .

\bibitem{knopp89}
G.~Knopp, A.~Loidl, K.~Knorr, L.~Pawlak, M.~Duczmal, R.~Caspary, U.~Gottwick,
  H.~Spille, F.~Steglich, and A.~P. Murani: Z. Phys. B
  {\bfseries 77} (1989) 95.

\bibitem{krimmel97}
A.~Krimmel, A.~Loidl, H.~Schober, and P.~C. Canfield: Phys. Rev. B {\bfseries
  55} (1997) 6416.

\bibitem{knopp87}
G.~Knopp, H.~Spille, A.~Loidl, K.~Knorr, U.~Rauchschwalbe, R.~Felten, G.~Weber,
  F.~Steglich, and A.~P. Murani: J. Magn. Magn. Mater.
  {\bfseries 63-64} (1987) 88.

\bibitem{boer87}
F.~R. de~Boer, J.~C.~P. Klaasse, P.~A. Veenhuizen, A.~B\"ohm, C.~D. Bredl,
  U.~Gottwick, H.~M. Mayer, L.~Pawlak, U.~Rauchschwalbe, H.~Spille, and
  F.~Steglich: J. Magn. Magn. Mater. {\bfseries 63-64}
  (1987) 91.

\bibitem{zwicknagl07}
G.~Zwicknagl: J. Low Temp. Phys. {\bfseries 147} (2007) 123.

\bibitem{jaccard92}
D.~Jaccard, K.~Behnia, and J.~Sierro: Phys. Lett. A {\bfseries 163} (1992)
  475.

\bibitem{jaccard99}
D.~Jaccard, H.~Wilhelm, K.~Alami-Yadri, and E.~Vargoz: Physica B
 {\bfseries 259-261} (1999) 1.

\bibitem{steglich79}
F.~Steglich, J.~Aarts, C.~D. Bredl, W.~Lieke, D.~Meschede, W.~Franz, and
  H.~Sch\"afer: Phys. Rev. Lett. {\bfseries 43} (1979) 1892.

\bibitem{nakamura88}
H.~Nakamura, Y.~Kitaoka, H.~Yamada, and K.~Asayama: J. Magn. Magn. Mater. 
{\bfseries 76-77} (1988) 517.

\bibitem{uemura88}
Y.~J. Uemura, W.~J. Kossler, X.~H. Yu, H.~E. Schone, J.~R. Kempton, C.~E.
  Stronach, S.~Barth, F.~N. Gygax, B.~Hitti, A.~Schenck, C.~Baines, W.~F.
  Lankford, Y.~Onuki, and T.~Komatsubara: Physica C
  {\bfseries 153-155} (1988) 455.

\bibitem{uemura89}
Y.~J. Uemura, W.~J. Kossler, X.~H. Yu, H.~E. Schone, J.~R. Kempton, C.~E.
  Stronach, S.~Barth, F.~N. Gygax, B.~Hitti, A.~Schenck, C.~Baines, W.~F.
  Lankford, Y.~\ifmmode~\bar{O}\else \={O}\fi{}nuki, and T.~Komatsubara: Phys.
  Rev. B {\bfseries 39} (1989) 4726.

\bibitem{gegenwart98}
P.~Gegenwart, C.~Langhammer, C.~Geibel, R.~Helfrich, M.~Lang, G.~Sparn,
  F.~Steglich, R.~Horn, L.~Donnevert, A.~Link, and W.~Assmus: Phys. Rev. Lett.
  {\bfseries 81} (1998) 1501.

\bibitem{steglich96}
F.~Steglich, P.~Gegenwart, C.~Geibel, R.~Helfrich, P.~Hellmann, M.~Lang,
  A.~Link, R.~Modler, G.~Sparn, N.~B{\"u}ttgen, and A.~Loidl: Physica B
   {\bfseries 223-224} (1996) 1.

\bibitem{stockert06b}
O.~Stockert, D.~Andreica, A.~Amato, H.~S. Jeevan, C.~Geibel, and F.~Steglich:
  Physica B {\bfseries 374-375} (2006) 167.

\bibitem{sparn98}
G.~Sparn, L.~Donnevert, P.~Hellmann, R.~Horn, F.~Laube, A.~Link, F.~Thomas,
  P.~Gegenwart, B.~Buschinger, C.~Geibel, and F.~Steglich: Rev. High Pressure
  Sci. Technol. {\bfseries 7} (1998) 431.

\bibitem{lengyel11}
E.~Lengyel, M.~Nicklas, H.~S. Jeevan, C.~Geibel, and F.~Steglich: Phys. Rev.
  Lett. {\bfseries 107} (2011) 057001.

\bibitem{knebel96}
G.~Knebel, C.~Eggert, D.~Engelmann, R.~Viana, A.~Krimmel, M.~Dressel, and
  A.~Loidl: Phys. Rev. B {\bfseries 53} (1996) 11586.

\bibitem{krimmel97a}
A.~Krimmel and A.~Loidl: Physica B {\bfseries 234-236} (1997)
  877.

\bibitem{stockert03}
O.~Stockert, M.~Deppe, C.~Geibel, F.~Steglich, D.~Hohlwein, and R.~Schneider:
  Acta Phys. Pol. B {\bfseries 34} (2003) 963.

\bibitem{faulhaber04}
E.~Faulhaber, O.~Stockert, M.~Rheinst\"adter, M.~Deppe, C.~Geibel,
  M.~Loewenhaupt, and F.~Steglich: J. Magn. Magn. Mater.
  {\bfseries 272-276} (2004) 44.

\bibitem{stockert05}
O.~Stockert, M.~Deppe, E.~Faulhaber, H.~Jeevan, R.~Schneider,
  N.~St{\"u}{\ss}er, C.~Geibel, M.~Loewenhaupt, and F.~Steglich: Physica B
   {\bfseries 359-361} (2005) 349.

\bibitem{faulhaber06}
E.~Faulhaber, O.~Stockert, B.~Grenier, B.~Ouladdiaf, M.~Deppe, C.~Geibel,
  F.~Steglich, and M.~Loewenhaupt: Physica B {\bfseries
  378-380} (2006) 78.

\bibitem{faulhaber08}
E.~Faulhaber: Dr. Thesis, Technische Universit\"at Dresden (2008).

\bibitem{stockert04}
O.~Stockert, E.~Faulhaber, G.~Zwicknagl, N.~St\"usser, H.~S. Jeevan, M.~Deppe,
  R.~Borth, R.~K\"uchler, M.~Loewenhaupt, C.~Geibel, and F.~Steglich: Phys. Rev. Lett.
   {\bfseries 92} (2004) 136401.

\bibitem{sparn06}
G.~Sparn, O.~Stockert, F.~Grosche, H.~Yuan, E.~Faulhaber, C.~Geibel, M.~Deppe,
  H.~Jeevan, M.~Loewenhaupt, G.~Zwicknagl, and F.~Steglich: J. Phys. Chem. Solids
   {\bfseries 67} (2006) 529.

\bibitem{stockert06a}
O.~Stockert, E.~Faulhaber, K.~Schmalzl, W.~Schmidt, H.~S. Jeevan, M.~Deppe,
  C.~Geibel, T.~Cichorek, T.~Nakanishi, M.~Loewenhaupt, and F.~Steglich:
  J. Phys.: Conf. Series {\bfseries 51} (2006) 211.

\bibitem{stockert07a}
O.~Stockert, M.~Nicklas, P.~Thalmeier, P.~Gegenwart, and F.~Steglich: {\em
  Handbook of Magnetism and Advanced Magnetic Materials}, ed. H.~Kronm\"uller
  and S.~Parkin (John Wiley \& Sons Ltd., 2007), Vol. 4: Novel Materials, Chap.
  Magnetism and Quantum Criticality in Heavy-Fermion Compounds: Interplay with
  Superconductivity, p. 2461.

\bibitem{arndt10}
J.~Arndt, O.~Stockert, E.~Faulhaber, P.~Fouquet, H.~S. Jeevan, C.~Geibel,
  M.~Loewenhaupt, and F.~Steglich: J. Phys.: Conf. Series
  {\bfseries 200} (2010) 012009 (4pp).

\bibitem{faulhaber07}
E.~Faulhaber, O.~Stockert, K.~Schmalzl, H.~S. Jeevan, M.~Deppe, C.~Geibel,
  F.~Steglich, and M.~Loewenhaupt: J. Magn. Magn. Mater.
  {\bfseries 310} (2007) 295.

\bibitem{stockert08}
O.~Stockert, J.~Arndt, A.~Schneidewind, H.~Schneider, H.~S. Jeevan, C.~Geibel,
  F.~Steglich, and M.~Loewenhaupt: Physica B {\bfseries 403}
  (2008) 973.

\bibitem{stockert11}
O.~Stockert, J.~Arndt, E.~Faulhaber, C.~Geibel, H.~S. Jeevan, S.~Kirchner,
  M.~Loewenhaupt, K.~Schmalzl, W.~Schmidt, Q.~Si, and F.~Steglich: Nat. Phys.
  {\bfseries 7} (2011) 119.

\bibitem{arndt11}
J.~Arndt, O.~Stockert, K.~Schmalzl, E.~Faulhaber, H.~S. Jeevan, C.~Geibel,
  W.~Schmidt, M.~Loewenhaupt, and F.~Steglich: Phys. Rev. Lett. {\bfseries 106}
  (2011) 246401.

\bibitem{ohkawa87}
F.~Ohkawa: J. Phys. Soc. Jpn. {\bfseries 56} (1987) 2267.

\bibitem{ishida99}
K.~Ishida, Y.~Kawasaki, K.~Tabuchi, K.~Kashima, Y.~Kitaoka, K.~Asayama,
  C.~Geibel, and F.~Steglich: Phys. Rev. Lett. {\bfseries 82} (1999)
  5353.

\bibitem{fujiwara08}
K.~Fujiwara, Y.~Hata, K.~Kobayashi, K.~Miyoshi, J.~Takeuchi, Y.~Shimaoka,
  H.~Kotegawa, T.~C. Kobayashi, C.~Geibel, and F.~Steglich: J. Phys. Soc. Jpn.
   {\bfseries 77} (2008) 123711.

\bibitem{rauchschwalbe82}
U.~Rauchschwalbe, W.~Lieke, C.~D. Bredl, F.~Steglich, J.~Aarts, K.~M. Martini,
  and A.~C. Mota: Phys. Rev. Lett. {\bfseries 49} (1982) 1448.

\bibitem{Kohn.65}
W.~Kohn and J.~M. Luttinger: Phys. Rev. Lett. {\bfseries 15} (1965) 524.

\bibitem{Onishi.00}
Y.~Onishi and K.~Miyake: J. Phys. Soc. Jpn. {\bfseries
  69} (2000) 3955.

\bibitem{Monthoux.04}
P.~Monthoux and G.~G. Lonzarich: Phys. Rev. B {\bfseries 69} (2004) 064517.

\bibitem{roehler88}
J.~R\"ohler, J.~Klug, and K.~Keulerz: J. Magn. Magn. Mater.
 {\bfseries 76-77} (1988) 340.

\bibitem{rueff11}
J.-P. Rueff, S.~Raymond, M.~Taguchi, M.~Sikora, J.-P. Iti\'e, F.~Baudelet,
  D.~Braithwaite, G.~Knebel, and D.~Jaccard: Phys. Rev. Lett. {\bfseries 106}
  (2011) 186405.

\bibitem{yuan06}
H.~Q. Yuan, F.~M. Grosche, M.~Deppe, G.~Sparn, C.~Geibel, and F.~Steglich:
  Phys. Rev. Lett. {\bfseries 96} (2006) 047008.

\bibitem{Holmes.07}
A.~T. Holmes, D.~Jaccard, and K.~Miyake: J. Phys. Soc. Jpn.
 {\bfseries 76} (2007) 051002.

\bibitem{onodera02}
A.~Onodera, S.~Tsuduki, Y.~Ohishi, T.~Watanuki, K.~Ishida, Y.~Kitaoka, and
  Y.~Onuki: Solid State Commun. {\bfseries 123} (2002) 113.

\bibitem{Watanabe.11}
S.~Watanabe and K.~Miyake: J. Phys.: Condens. Matter {\bfseries 23}
  (2011) 094217.

\bibitem{lengyel09}
E.~Lengyel, M.~Nicklas, H.~S. Jeevan, G.~Sparn, C.~Geibel, F.~Steglich,
  Y.~Yoshioka, and K.~Miyake: Phys. Rev. B {\bfseries 80} (2009) 140513.

\bibitem{vieyra11}
H.~A. Vieyra, N.~Oeschler, S.~Seiro, H.~S. Jeevan, C.~Geibel, D.~Parker, and
  F.~Steglich: Phys. Rev. Lett. {\bfseries 106} (2011) 207001.

\bibitem{stock08}
C.~Stock, C.~Broholm, J.~Hudis, H.~J. Kang, and C.~Petrovic: Phys. Rev. Lett.
 {\bfseries 100} (2008) 087001.

\bibitem{eremin08}
I.~Eremin, G.~Zwicknagl, P.~Thalmeier, and P.~Fulde: Phys. Rev. Lett.
  {\bfseries 101} (2008) 187001.

\bibitem{Chubukov.08}
A.~V. Chubukov and L.~P. Gor'kov: Phys. Rev. Lett. {\bfseries 101} (2008)
  147004.

\bibitem{schlabitz84}
W.~Schlabitz and {et al.}: Abstract, ICVF, Cologne (1984).

\bibitem{schlabitz86}
W.~Schlabitz, J.~Baumann, B.~Pollit, U.~Rauchschwalbe, H.~M. Mayer, U.~Ahlheim,
  and C.~D. Bredl: Z. Phys. B {\bfseries 62}
  (1986) 171.

\bibitem{palstra85}
T.~T.~M. Palstra, A.~A. Menovsky, J.~v.~d. Berg, A.~J. Dirkmaat, P.~H. Kes,
  G.~J. Nieuwenhuys, and J.~A. Mydosh: Phys. Rev. Lett. {\bfseries 55} (1985)
  2727.

\bibitem{maple86}
M.~B. Maple, J.~W. Chen, Y.~Dalichaouch, T.~Kohara, C.~Rossel, M.~S.
  Torikachvili, M.~W. McElfresh, and J.~D. Thompson: Phys. Rev. Lett.
  {\bfseries 56} (1986) 185.

\bibitem{broholm87}
C.~Broholm, J.~K. Kjems, W.~J.~L. Buyers, P.~Matthews, T.~T.~M. Palstra, A.~A.
  Menovsky, and J.~A. Mydosh: Phys. Rev. Lett. {\bfseries 58} (1987) 1467.

\bibitem{barzykin95}
V.~Barzykin and L.~P. Gor'kov: Phys. Rev. Lett. {\bfseries 74} (1995) 4301.

\bibitem{santini94}
P.~Santini and G.~Amoretti: Phys. Rev. Lett. {\bfseries 73} (1994) 1027.

\bibitem{ohkawa99}
F.~J. Ohkawa and H.~Shimizu: J. Phys.: Condens. Matter {\bfseries
  11} (1999) L519.

\bibitem{kuramoto06}
Y.~Kuramoto, H.~Kusunose, and A.~Kiss: Physica B {\bfseries
  383} (2006) 5.

\bibitem{chandra02}
P.~Chandra, P.~Coleman, J.~A. Mydosh, and V.~Tripathi: Nature {\bfseries 417}
  (2002) 831.

\bibitem{mineev05}
V.~P. Mineev and M.~E. Zhitomirsky: Phys. Rev. B {\bfseries 72} (2005) 014432.

\bibitem{varma06}
C.~M. Varma and L.~Zhu: Phys. Rev. Lett. {\bfseries 96} (2006) 036405.

\bibitem{wiebe07}
C.~R. Wiebe, J.~A. Janik, G.~J. MacDougall, G.~M. Luke, J.~D. Garrett, H.~D.
  Zhou, Y.~J. Jo, L.~Balicas, Y.~Qiu, J.~R.~D. Copley, Z.~Yamani, and W.~J.~L.
  Buyers: Nat. Phys. {\bfseries 3} (2007) 96.

\bibitem{fawcett88}
E.~Fawcett: Rev. Mod. Phys. {\bfseries 60} (1988) 209.

\bibitem{santander-syro09}
A.~F. Santander-Syro, M.~Klein, F.~L. Boariu, A.~Nuber, P.~Lejay, and
  F.~Reinert: Nat. Phys. {\bfseries 5} (2009) 637.

\bibitem{schmidt10}
A.~R. Schmidt, M.~H. Hamidian, P.~Wahl, F.~Meier, A.~V. Balatsky, J.~D.
  Garrett, T.~J. Williams, G.~M. Luke, and J.~C. Davis: Nature {\bfseries 465}
  (2010) 570.

\bibitem{aynajian10}
P.~Aynajian, E.~H. da~Silva~Neto, C.~V. Parker, Y.~Huang, A.~Pasupathy,
  J.~Mydosh, and A.~Yazdani: Proc. Natl. Acad. Sci. USA
  {\bfseries 107} (2010) 10383.

\bibitem{martin82}
R.~M. Martin: Phys. Rev. Lett. {\bfseries 48} (1982) 362.

\bibitem{kim90}
C.-I. Kim, Y.~Kuramoto, and T.~Kasuya: J. Phys. Soc. Jpn.
  {\bfseries 59} (1990) 2414.

\bibitem{okazaki11}
R.~Okazaki, T.~Shibauchi, H.~J. Shi, Y.~Haga, T.~D. Matsuda, E.~Yamamoto,
  Y.~Onuki, H.~Ikeda, and Y.~Matsuda: Science {\bfseries 331} (2011) 439.

\bibitem{Borzi.07}
R.~A. Borzi, S.~A. Grigera, J.~Farrell, R.~S. Perry, S.~J.~S. Lister, S.~L.
  Lee, D.~A. Tennant, Y.~Maeno, and A.~P. Mackenzie: Science {\bfseries 315}
  (2007) 214.

\bibitem{haule09}
K.~Haule and G.~Kotliar: Nat. Phys. {\bfseries 5} (2009) 796.

\bibitem{kohara86}
T.~Kohara, Y.~Kohori, K.~Asayama, Y.~Kitaoka, M.~B. Maple, and M.~S.
  Torikachvili: Solid State Commun. {\bfseries 59} (1986) 603.

\bibitem{cox98}
D.~L. Cox and A.~Zawadowski: Adv. Phys. {\bfseries 47} (1998) 599.

\bibitem{fujimori11}
S.-I. Fujimori and {et al.}: J. Phys. Soc. Jpn., in press.

\bibitem{amitsuka07}
H.~Amitsuka, K.~Matsuda, I.~Kawasaki, K.~Tenya, M.~Yokoyama, C.~Sekine,
  N.~Tateiwa, T.~Kobayashi, S.~Kawarazaki, and H.~Yoshizawa: J. Magn. Magn. Mater.
   {\bfseries 310} (2007) 214.

\bibitem{hassinger08}
E.~Hassinger, G.~Knebel, K.~Izawa, P.~Lejay, B.~Salce, and J.~Flouquet: Phys.
  Rev. B {\bfseries 77} (2008) 115117.

\bibitem{Kasahara.07}
Y.~Kasahara, T.~Iwasawa, H.~Shishido, T.~Shibauchi, K.~Behnia, Y.~Haga, T.~D.
  Matsuda, Y.~Onuki, M.~Sigrist, and Y.~Matsuda: Phys. Rev. Lett. {\bfseries
  99} (2007) 116402.

\bibitem{Yano.08}
K.~Yano, T.~Sakakibara, T.~Tayama, M.~Yokoyama, H.~Amitsuka, Y.~Homma,
  P.~Miranovi\ifmmode~\acute{c}\else \'{c}\fi{}, M.~Ichioka, Y.~Tsutsumi, and
  K.~Machida: Phys. Rev. Lett. {\bfseries 100} (2008) 017004.

\bibitem{hertz76}
J.~A. Hertz: Phys. Rev. B {\bfseries 14} (1976) 1165.

\bibitem{Moriya}
T.~Moriya: {\em Spin Fluctuations in Itinerant Electron Magnetism} (Springer,
  Berlin, 1985).

\bibitem{millis93}
A.~J. Millis: Phys. Rev. B {\bfseries 48} (1993) 7183.

\bibitem{si01}
Q.~Si, S.~Rabello, K.~Ingersent, and J.~L. Smith: Nature {\bfseries 413} (2001)
  804.

\bibitem{coleman01}
P.~Coleman, C.~P\'epin, Q.~Si, and R.~Ramazashvili: J. Phys.: Condens. Matter
 {\bfseries 13} (2001) R723.

\bibitem{senthil04}
T.~Senthil, M.~Vojta, and S.~Sachdev: Phys. Rev. B {\bfseries 69} (2004)
  035111.

\bibitem{paul07}
I.~Paul, C.~P\'epin, and M.~R. Norman: Phys. Rev. Lett. {\bfseries 98} (2007)
  026402.

\bibitem{schmitt-rink86}
S.~Schmitt-Rink, K.~Miyake, and C.~M. Varma: Phys. Rev. Lett. {\bfseries 57}
  (1986) 2575.

\bibitem{scalapino86}
D.~J. Scalapino, E.~Loh, and J.~E. Hirsch: Phys. Rev. B {\bfseries 34} (1986)
  8190.

\bibitem{scalapino98}
D.~J. Scalapino and S.~R. White: Phys. Rev. B {\bfseries 58} (1998) 8222.

\bibitem{leggett98}
A.~J. Leggett: J. Phys. Chem. Solids {\bfseries 59} (1998)
  1729.

\bibitem{shishido05}
H.~Shishido, R.~Settai, H.~Harima, and Y.~\={O}nuki: J. Phys. Soc. Jpn.
 {\bfseries 74} (2005) 1103.

\bibitem{knebel08}
G.~Knebel, D.~Aoki, J.-P. Brison, and J.~Flouquet: J. Phys. Soc. Jpn.
 {\bfseries 77} (2008) 114704.

\bibitem{grewe91}
N.~Grewe and F.~Steglich: {\em Handbook on the Physics and Chemistry of Rare
  Earths}, ed. K.~A. {Gschneidner, Jr.} and L.~Eyring (Elsevier, 1991),
  Vol.~14, Chap. 97, Heavy Fermions, p. 343.

\bibitem{Ernst.11}
S.~Ernst, S.~Kirchner, C.~Krellner, C.~Geibel, G.~Zwicknagl, F.~Steglich, and
  S.~Wirth: Nature {\bfseries 474} (2011) 362.

\bibitem{Kamihara.08}
Y.~Kamihara, T.~Watanabe, M.~Hirano, and H.~Hosono: J. Am.
  Chem. Soc. {\bfseries 130} (2008) 3296.

\bibitem{ishida09}
K.~Ishida, Y.~Nakai, and H.~Hosono: J. Phys. Soc. Jpn.
  {\bfseries 78} (2009) 062001.

\bibitem{qazilbash09}
M.~M. Qazilbash, J.~J. Hamlin, R.~E. Baumbach, L.~Zhang, D.~J. Singh, M.~B.
  Maple, and D.~N. Basov: Nat. Phys. {\bfseries 5} (2009) 647.

\bibitem{fang11}
M.-H. Fang, H.-D. Wang, C.-H. Dong, Z.-J. Li, C.-M. Feng, J.~Chen, and H.~Q.
  Yuan: Europhys. Lett. {\bfseries 94} (2011) 27009.

\end{thebibliography}

\end{document}